\documentclass[prb,aps,twocolumn,showpacs,floatfix,superscriptaddress]{revtex4}
\usepackage{graphicx}
\usepackage{bm}
\usepackage{amsmath,amsfonts}
\usepackage{amsbsy}

\begin{document}

\title{Scattering of charge and spin excitations and equilibration of a one-dimensional
  Wigner crystal}

\author{K. A. Matveev}

\affiliation{Materials Science Division, Argonne National Laboratory,
  Argonne, Illinois 60439, USA}

\author{A. V. Andreev}

\affiliation{Department of Physics, University of Washington, Seattle,
  Washington 98195, USA}

\author{A. D. Klironomos}

\affiliation{American Physical Society, 1 Research Road, Ridge, New York 11961-9000}

\date{July 28, 2014}

\begin{abstract}

  We study scattering of charge and spin excitations in a system of
  interacting electrons in one dimension.  At low densities electrons
  form a one-dimensional Wigner crystal.  To a first approximation the
  charge excitations are the phonons in the Wigner crystal, and the
  spin excitations are described by the Heisenberg model with nearest
  neighbor exchange coupling.  This model is integrable and thus
  incapable of describing some important phenomena, such as scattering
  of excitations off each other and the resulting equilibration of the
  system.  We obtain the leading corrections to this model, including
  charge-spin coupling and the next-nearest neighbor exchange in the
  spin subsystem.  We apply the results to the problem of
  equilibration of the one-dimensional Wigner crystal and find that
  the leading contribution to the equilibration rate arises from
  scattering of spin excitations off each other.  We discuss the
  implications of our results for the conductance of quantum wires at
  low electron densities.
\end{abstract}

\pacs{71.10.Pm}

\maketitle

\section{Introduction}\label{sec:intro}

The low-energy properties of one-dimensional electron systems are
commonly described in the framework of the Tomonaga-Luttinger liquid
theory.  \cite{giamarchi2004quantum} In this approach the electrons
are described in terms of elementary excitations with bosonic
statistics, which have the meaning of waves of charge and spin
densities.  These waves propagate at different velocities,
\cite{dzyaloshinskii_correlation_1974} resulting in the separation of
the charge and spin of the electrons.  A detailed theory of
spin-charge separation can be developed in the case of strong
repulsive interactions.  This was first accomplished for the
one-dimensional Hubbard model by Ogata and
Shiba,\cite{ogata_bethe-ansatz_1990} who showed that the ground state
wave function of the system can be expressed as a product of two
separate wave functions describing the charge and spin degrees of
freedom.

Experimentally, one-dimensional electron systems are often realized in
GaAs quantum wires.\cite{berggren_new_2002} In contrast to the Hubbard
model, electrons in quantum wires are not confined to discrete lattice
sites, and interact via the long-range Coulomb repulsion
$V(x)=e^2/\varepsilon |x|$.  The density of electrons in these systems
is easily controlled by gates.  At low density $n\ll a_B^{-1}$ the
Coulomb repulsion between electrons is much larger than their kinetic
energy, and in the ground state the system forms a Wigner
crystal.\cite{wigner_interaction_1934} (Here
$a_B=\varepsilon\hbar^2/me^2$ is the Bohr radius in the material,
$\varepsilon$ is the dielectric constant, and $m$ is the effective
mass of the electrons.)

This physical picture of strongly interacting one-dimensional electron
systems enables a simple description of the charge excitations as
phonons in the Wigner crystal.  Mathematically, they are accounted for
by the phonon Hamiltonian
\begin{equation}
  \label{eq:H_rho}
  H_\rho^{(0)}=\sum_l \frac{p_l^2}{2m}
        +\frac{e^2}{2\varepsilon a^3}
         \sum_{l\neq l'}\frac{(u_l-u_{l'})^2}{|l-l'|^3},
\end{equation}
where $a=n^{-1}$ is the average interelectron distance, and the $l$th
electron is described by its momentum $p_l$ and displacement from the
equilibrium position $u_l=x_l-la$.

As long as the displacements $u_l$ are small, the spins are attached
to the lattice sites.  In order for the spins to move along the
crystal, neighboring electrons must be able to switch places on the
Wigner lattice.  Such processes lead to the exchange coupling of the
spins\cite{matveev_conductance_2004}
\begin{equation}
  \label{eq:spin-chain}
  H_\sigma^{(0)} = \sum_l J\, {\bm S}_{l}\cdot{\bm S}_{l+1}.
\end{equation}
Because the exchange process involves two adjacent electrons tunneling
through the strong Coulomb barrier
\mbox{$e^2/\varepsilon|x_l-x_{l+1}|$}, the coupling constant $J$ is
exponentially small
\begin{equation}
  \label{eq:J}
  J \sim (na_B)^{5/4}\frac{e^2}{\varepsilon a_B}
    \exp\left(-\frac{\eta}{\sqrt{na_B}}\right),
\end{equation}
where $\eta\approx 2.798$.\cite{matveev_conductance_2004,
  klironomos_exchange_2005, fogler_exchange_2005}

The Hamiltonian given by Eqs.~(\ref{eq:H_rho}) and
(\ref{eq:spin-chain}) describes the charge and spin excitations in the
Wigner crystal near the ground state.  At finite temperature the
electron density fluctuates, resulting in thermal fluctuations of
the exchange constant (\ref{eq:J}).  This limits the applicability of
the Hamiltonian (\ref{eq:H_rho}) and (\ref{eq:spin-chain}) to
relatively low temperatures, $T\ll(e^2/\varepsilon a_B)(na_B)^{7/4}$.
Note that this range includes the most interesting temperature
regime $T\lesssim J$.

As expected, the charge and spin fluctuations described by
Eqs.~(\ref{eq:H_rho}) and (\ref{eq:spin-chain}) are decoupled from
each other.  Another interesting feature of this Hamiltonian is its
integrability.  Indeed, Eq.~(\ref{eq:H_rho}) describes noninteracting
phonons, and integrability of the spin-$\frac12$ Heisenberg model
(\ref{eq:spin-chain}) was shown by Bethe \cite{bethe_zur_1931} in
1931.  A defining characteristic of integrable models is a large
number of integrals of motion, which prevents relaxation of the system
to equilibrium.  Thus, although the Hamiltonian (\ref{eq:H_rho}),
(\ref{eq:spin-chain}) gives an adequate description of the equilibrium
properties of the system, corrections to it must be considered in
order to discuss the approach to equilibrium.  Examples of physical
effects controlled by equilibration include various transport
phenomena in high-mobility quantum
wires.\cite{micklitz_transport_2010, levchenko_transport_2010,
  andreev_hydrodynamic_2011, matveev_equilibration_2011,
  levchenko_interaction_2011, dmitriev_coulomb_2012,
  apostolov_thermal_2013}

The first goal of this paper is to identify and evaluate the leading
order corrections to the Hamiltonian (\ref{eq:H_rho}),
(\ref{eq:spin-chain}).  Such corrections fall into three categories.
First, one should note that the Wigner crystal is not a perfectly
harmonic chain.  Anharmonic corrections to the Hamiltonian
(\ref{eq:H_rho}) are determined by the third and higher derivatives of
the interaction potential and are well understood.  Corrections of the
second type account for coupling between the charge and spin
excitations.  They can be understood by noting that the exchange
constant $J$ in Eq.~(\ref{eq:spin-chain}) depends on the electron
density, Eq.~(\ref{eq:J}), which fluctuates as the phonons propagate
through the system.  Corrections of the third type appear in the spin
channel.  Similarly to the nearest neighbor exchange coupling
(\ref{eq:spin-chain}), they are caused by quantum tunneling processes
resulting in some electrons changing places on the Wigner lattice.

Our second goal is to develop a theory of equilibration of a
one-dimensional Wigner crystal at the lowest temperatures $T\ll J$.
The full equilibration of one-dimensional systems is an exponentially
slow process, as it requires backscattering of highly excited hole
quasiparticles.\cite{micklitz_transport_2010, matveev_equilibration_2010,
  matveev_equilibration_2012, matveev_scattering_2012} In the case of
strong interactions the hole becomes a spinon excitation in the spin
chain formed on the lattice sites of the Wigner crystal, see
Eq.~(\ref{eq:spin-chain}).  Because the typical energy of the spinon
$J$ is small compared to the Fermi energy, the equilibration rate is
greatly enhanced in this regime.  It is dominated by the processes of
scattering of spinons by the thermal excitations of the system.  The
evaluation of the scattering rate relies heavily on the preceding
results for the form and magnitude of the integrability-breaking
perturbations in the one-dimensional Wigner crystal.

We discuss the corrections to the Hamiltonian (\ref{eq:H_rho}),
(\ref{eq:spin-chain}) in Sec~\ref{sec:corrections}.  The study of the
exchange processes beyond the nearest neighbor coupling requires the
calculation of the relevant tunneling amplitudes in the WKB
approximation.  We evaluate the WKB action for the dominant
three-particle exchange process in Sec.~\ref{sec:WKB}.  In addition to
corrections to the nearest neighbor exchange constant (\ref{eq:J}), this
process generates exchange coupling of the next nearest neighbor spins
in the crystal, which breaks the integrability of the Heisenberg chain.
In Sec.~\ref{sec:rate-full-equil} we use the results of
Secs.~\ref{sec:corrections} and \ref{sec:WKB} to evaluate the rate of
full equilibration of the Wigner crystal at low temperatures.  In
Sec.~\ref{sec:summary} we summarize our results and discuss their
implications for the temperature dependence of conductance of quantum
wires.

\section{Hamiltonian of the Wigner crystal}
\label{sec:corrections}

One-dimensional electrons interacting via sufficiently strong
repulsion form a periodic chain regardless of the spatial dependence
of the interaction potential $V(x)$, except for an extremely short-ranged
interaction.  In this paper we focus on the case of pure Coulomb
repulsion $V(x)=e^2/\varepsilon|x|$, corresponding to the traditional
definition of the Wigner crystal.  The regime of strong repulsion is
realized at low electron density $n$, when the typical energy of
Coulomb repulsion $(e^2/\varepsilon) n$ greatly exceeds the typical
kinetic energy $(\hbar^2/m)n^2$ in a free electron gas.  Thus the
Wigner crystal limit is achieved at $na_B\ll1$, or at
\begin{equation}
  \label{eq:r_s}
  r_s=\frac{1}{2na_B}\gg1.
\end{equation}
It is worth mentioning that in one dimension quantum fluctuations
destroy long-range order even at zero temperature, and the Wigner
crystal picture refers only to the short-range ordering of electrons.

The full microscopic Hamiltonian of one-dimensional electrons with
Coulomb interactions is given by
\begin{equation}
  \label{eq:microscopic_Hamiltonian}
  H=\sum_l \frac{p_l^2}{2m}
    +\frac12\sum_{l\neq l'}\frac{e^2}{\varepsilon |x_l-x_{l'}|}.
\end{equation}
Our goal in this section is to develop the low energy description of
the system at $na_B\ll1$.

\subsection{Spin-charge separation}
\label{sec:spin-charge-separation}

To leading order in $na_B$, electrons do not switch places on the
Wigner lattice.  Indeed, in order to do so one-dimensional electrons
must approach each other and experience strong Coulomb repulsion.  As
a result, this process is essentially tunneling under the Coulomb
barrier,\footnote{Strictly speaking, the one-dimensional Coulomb
  barrier is impenetrable.\cite{andrews_singular_1976} On the other
  hand, quantum wires always have finite width $w$, which results in
  finite tunneling amplitude depending only logarithmically on $w$,
  see Ref.~\onlinecite{fogler_exchange_2005}.} and its amplitude is
exponentially small.  As long as electrons do not switch places, their
spins do not interact, and each state of $N$ particles is $2^N$-fold
degenerate.

The effect of the tunneling processes can be understood as
follows. Because the Hamiltonian (\ref{eq:microscopic_Hamiltonian})
does not depend on spins, the eigenstates of the system factorize into
the product of the coordinate and spin components:
\begin{equation}
  \label{eq:factorization}
  \psi(x_1,\sigma_1; \ldots;x_N,\sigma_N)
    =\chi_{\sigma_1,\ldots,\sigma_N}\phi(x_1,\ldots,x_N).
\end{equation}
The coordinate wavefunction $\phi(x_1,\ldots,x_N)$ satisfies
\begin{equation}
  \label{eq:Schroedinger}
  H\phi=E\phi
\end{equation}
with $H$ given by Eq.~(\ref{eq:microscopic_Hamiltonian}).  One can
interpret Eq.~(\ref{eq:Schroedinger}) as  the
Schr\"odinger equation for a system of $N$ spinless distinguishable
particles.
In the absence of tunneling the spatial
ordering of the particles is preserved.  Furthermore, for any state
$\phi(x_1,\ldots,x_N)$, there are a total of $N!$ degenerate states
obtained from it by all permutations of the coordinates $x_1$, \ldots,
$x_N$.  The strongest tunneling process permutes two adjacent
particles.  Thus, at low energies, the Hamiltonian
(\ref{eq:microscopic_Hamiltonian}) can be approximated as
\begin{equation}
  \label{eq:H_approximated}
  H=H_\rho-\frac{J}{2}\sum_l P(x_l,x_{l+1}),
\end{equation}
where $P(x_i,x_j)$ is the operator of the permutation of the
coordinates of the $i$th and $j$th particles.  The constant $J$ can be
computed in the WKB approximation\cite{matveev_conductance_2004,
  klironomos_exchange_2005, fogler_exchange_2005} and is given by
Eq.~(\ref{eq:J}).

The operator $H_\rho$ in Eq.~(\ref{eq:H_approximated}) coincides with
Eq.~(\ref{eq:microscopic_Hamiltonian}) with the additional condition
that no switching of the position of the particles is allowed.  It can
be formally defined by adding to
Eq.~(\ref{eq:microscopic_Hamiltonian}) the point-like repulsive
potential $A\delta(x_l-x_{l'})$ with $A\to+\infty$.  In the Wigner
crystal approximation one expresses the coordinates of the particles
in terms of their displacements from lattice sites, $x_l=la+u_l$.
Expanding the interaction term to second order in $u_l$ one obtains
the phonon Hamiltonian (\ref{eq:H_rho}).

When applying the result (\ref{eq:H_approximated}) to a system of
identical fermions one should bear in mind that the wavefunction
(\ref{eq:factorization}) changes sign upon simultaneous permutation
of coordinates $x_l\leftrightarrow x_{l+1}$ and spins
$\sigma_l\leftrightarrow \sigma_{l+1}$.  Thus for fermions the two
permutations are related as
$P(x_l,x_{l+1})=-P_{\sigma_l,\sigma_{l+1}}$, where
\begin{equation}
  \label{eq:spin_permutation}
  P_{\sigma_i,\sigma_{j}}= \frac12+ 2{\bm S}_{i}\cdot{\bm S}_{j}
\end{equation}
is the operator of permutation of spins $i$ and $j$.  Using this
result we find that the second term in Eq.~(\ref{eq:H_approximated})
gives the spin Hamiltonian (\ref{eq:spin-chain}).

Equations (\ref{eq:H_rho}) and (\ref{eq:spin-chain}) represent the
leading contributions to the low-energy Hamiltonian of the system.  As
discussed above, the resulting theory is integrable, which precludes
scattering of either charge or spin excitations.  The latter is made
possible by the subleading contributions to the Hamiltonian.  They
include corrections in the charge sector, in the spin sector, and the
coupling between the charge and spin sectors.

\subsection{Charge sector}
\label{sec:phonons}

It is convenient to rewrite Eq.~(\ref{eq:H_rho}) in terms of the
phonon operators $b_q$, $b_q^\dagger$ using the standard relations
\begin{eqnarray}
  \label{eq:second_quantization}
  u_l &=& \sum_q \sqrt{\frac{\hbar}{2mN\omega_q}}\, (b_q+b_{-q}^\dagger)
  e^{iql},
\\
  p_l &=& -i\sum_q \sqrt{\frac{\hbar m\omega_q}{2N}}\, (b_q-b_{-q}^\dagger)
  e^{iql}.
\end{eqnarray}
Here $N=nL$ is the total number of electrons in a system of size
$L$, and the phonon frequencies are found by solving the classical
equations of motion with the Hamiltonian (\ref{eq:H_rho}),
\begin{equation}
  \label{eq:omega_q}
  \hbar\omega_q=2(na_B)^{3/2}
                \frac{e^2}{\varepsilon a_B}
                \left[
                   \sum_{l=1}^\infty \frac{1-\cos (ql)}{l^3}
                \right]^{1/2}.
\end{equation}
This yields
\begin{equation}
  \label{eq:H_rho_phonons}
  H_\rho^{(0)}=\sum_q\hbar\omega_q\! \left(b_q^\dagger b_q^{}+\frac12\right).
\end{equation}
The behavior of $\omega_q$ at $q\to0$ is discussed in
Appendix~\ref{sec:frequencies}.

Beyond the harmonic approximation, one finds corrections to the
Hamiltonian (\ref{eq:H_rho_phonons}), which include terms of third and
higher powers in the phonon operators $b_q$ and $b_q^\dagger$ and can be
interpreted as interactions of the phonons.  Expanding the Coulomb
interaction in Eq.~(\ref{eq:microscopic_Hamiltonian}) to all orders in
electron displacements, one can express the anharmonic corrections to
$H_\rho^{(0)}$ as
\begin{eqnarray}
  H_\rho&=&H_\rho^{(0)}+\sum_{r=3}^\infty \hat V_\rho^{(r)},
\\
  \hat V_\rho^{(r)}&=&\frac{e^2}{\varepsilon a^{r+1}}
              \sum_{l>l'}\frac{(u_{l'}-u_l)^r}{(l-l')^{r+1}}.
  \label{eq:perturbation}
\end{eqnarray}
This expression can be equivalently written in terms of the phonon
operators using Eq.~(\ref{eq:second_quantization}).

\subsection{Spin sector}
\label{sec:spin-sector}

The Hamiltonian (\ref{eq:spin-chain}) accounts for the processes of
two nearest neighbor electrons switching positions on the Wigner
lattice.  Additional contributions to the spin Hamiltonian will appear
from any cyclic exchange process, including exchanges of two next-nearest neighbors, three consecutive electrons, four consecutive
electrons, etc.  The four processes we listed were considered in the
case of a quasi-one-dimensional Wigner
crystal.\cite{klironomos_spontaneous_2006, klironomos_spin_2007} The
range of electron densities considered there was above the transition
from a purely one-dimensional Wigner crystal into the zigzag phase.
Based on the results of Refs.~\onlinecite{klironomos_spontaneous_2006}
and \onlinecite{klironomos_spin_2007} in the region of lowest
densities studied, where the zigzag distortion is small, we conclude
that the strongest exchange is that of nearest neighbors, and the next
strongest one involves three consecutive electrons.

One can account for the three-particle exchange processes following
the prescription of Sec.~\ref{sec:spin-charge-separation}.  In
addition to the two-particle exchange term in
Eq.~(\ref{eq:H_approximated}) one obtains the contribution of the form
\begin{equation}
  \label{eq:three-particle_permutation}
  -\frac{\tilde J}{2}\sum_l [P(x_l,x_{l+1},x_{l+2})+P(x_l,x_{l+2},x_{l+1})],
\end{equation}
where the operator $P(x_i,x_j,x_k)$ performs a cyclic permutation of
coordinates $x_i\to x_j\to x_k\to x_i$.  Because this is an even
permutation, its outcome is equivalent to the permutation of electron
spins $P_{\sigma_i,\sigma_k,\sigma_j}$.  One can therefore rewrite the
perturbation (\ref{eq:three-particle_permutation}) as
\begin{eqnarray*}
  &&-\frac{\tilde J}{2}\sum_l
  [P_{\sigma_l,\sigma_{l+1},\sigma_{l+2}}+P_{\sigma_l,\sigma_{l+2},\sigma_{l+1}}]
\nonumber\\
  &&=   -\tilde J\sum_l
  \left[
   \frac14 + {\bm S}_{l}\cdot{\bm S}_{l+1}
           + {\bm S}_{l+1}\cdot{\bm S}_{l+2}
           + {\bm S}_{l}\cdot{\bm S}_{l+2}
  \label{eq:three-spin_permutation}
  \right],
\end{eqnarray*}
where we used the expression for the three-spin permutation in terms
of the spin operators obtained in
Ref.~\onlinecite{thouless_exchange_1965}.  Omitting the inessential
constant, we get the following perturbation in the spin Hamiltonian
due to the exchange of three consecutive electrons
\begin{equation}
  \label{eq:three-particle_perturbation}
  \hat V_\sigma = -\sum_l \tilde J\,
                [2\,{\bm S}_{l}\cdot{\bm S}_{l+1}
                 + {\bm S}_{l}\cdot{\bm S}_{l+2}].
\end{equation}

The constant $\tilde J$ will be evaluated in the WKB approximation in
Sec.~\ref{sec:WKB}.  We will see that $\tilde J$ is exponentially
smaller than the nearest neighbor exchange constant $J$.  Thus the
first term in the right-hand side of
Eq.~(\ref{eq:three-particle_perturbation}) gives only a negligible
correction to $J$.  On the other hand, the second term couples next
nearest neighbor spins.  Such perturbations break integrability of the
Heisenberg model (\ref{eq:spin-chain}) and give rise to relaxation
of spin excitations.

\subsection{Spin-charge coupling}
\label{sec:coupling-charge-spin}

So far we discussed corrections to the low-energy Hamiltonian of the
Wigner crystal given by Eqs.~(\ref{eq:H_rho}) and
(\ref{eq:spin-chain}), which give rise to the scattering of
excitations separately in the charge and spin sectors.  As a result of
such scattering the excitations equilibrate within each sector, but
not with the excitations in the other sector.  Full equilibration of
the system requires coupling of the charge and spin excitations, which
we discuss below.

The magnitude of the exchange constant $J$ depends strongly on the
electron density [see Eq.~(\ref{eq:J})].  The latter varies when
phonons propagate through the system, giving rise to the coupling of
the charge and spin degrees of freedom.\cite{thouless_exchange_1965}
One can account for this effect by evaluating $J$ in
Eq.~(\ref{eq:spin-chain}), using instead of the average density $n$ in
Eq.~(\ref{eq:J}) the position-dependent electron density in the
presence of the phonon deformation:
\begin{equation}
  \label{eq:density_operator}
  n\to \frac{1}{a+u_{l+1}-u_l}=\frac{n}{1+n(u_{l+1}-u_l)}.
\end{equation}
To lowest order in the phonon displacements and $na_B$, this procedure
results in the coupling of the charge and spin degrees of freedom in
the form
\begin{equation}
  \label{eq:v_rho_sigma_1}
  \hat V_{\rho\sigma} = -J\frac{\eta}{2\sqrt{na_B}}
                     \sum_l n(u_{l+1}-u_l)\,{\bm S}_{l}\cdot{\bm S}_{l+1}.
\end{equation}
The displacements $u_l$ in (\ref{eq:v_rho_sigma_1}) can be expressed
in terms of the phonon operators using
Eq.~(\ref{eq:second_quantization}).  The above procedure is justified
as long as the phonon density fluctuations remain approximately
uniform at the scale of interparticle distance, i.e., for phonons with
$|q|\ll1$.  The same condition ensures that the density fluctuations
are static on the scale of the WKB tunneling time.

\section{Three-particle exchange in the WKB approximation}
\label{sec:WKB}

Spin exchange in the Wigner crystal is caused by quantum tunneling
processes which allow electrons to switch places on the Wigner
lattice.  We discussed these exchange processes in
Secs.~\ref{sec:spin-charge-separation} and \ref{sec:spin-sector}.  In
particular, we found the expression
(\ref{eq:three-particle_perturbation}) for the leading order
correction to the spin Hamiltonian (\ref{eq:spin-chain}).  In this
section we discuss the magnitude of the respective exchange constant
$\tilde J$.

At small $na_B$, the tunneling amplitudes are small and can be
evaluated in the WKB approximation.  We perform the calculation using
the instanton technique.  In this approach the tunneling amplitudes
are controlled by the imaginary-time action
\begin{equation}
  \label{eq:initial_action}
  S[\{x_l(t)\}]=\int
     \left(
       \sum_l \frac{m}{2}\, \dot x_l^2
       +\sum_{l<l'} \frac{e^2}{\varepsilon|x_l-x_{l'}|}
     \right)dt.
\end{equation}
It is convenient to introduce the dimensionless coordinates $X_l$ and
time $\tau$ as
\begin{equation}
  \label{eq:rescaled_variables}
  x_l=\frac{1}{n}\, X_l,
  \quad
  t=\frac{\hbar\varepsilon a_B}{(na_B)^{3/2}e^2}\,\tau.
\end{equation}
This procedure brings Eq.~(\ref{eq:initial_action}) to the form
\begin{equation}
  \label{eq:action_rescaled}
  \frac1\hbar S[\{x_l(t)\}]=\frac{1}{\sqrt{na_B}}\,\eta[\{X_l(\tau)\}],
\end{equation}
where $\eta$ is the dimensionless action
\begin{equation}
  \label{eq:eta_definition}
  \eta[\{X_l(\tau)\}]=\int
     \left(
       \sum_l \frac{1}{2}\, \dot X_l^2
       +\sum_{l<l'} \frac{1}{|X_l-X_{l'}|}
     \right)d\tau.
\end{equation}
It is minimized for the trivial trajectory $X_l(\tau)=l$.

The cyclic exchange of three particles in
Eq.~(\ref{eq:three-particle_permutation}) is described by trajectories
with boundary conditions $X_l=l$ at $\tau\to -\infty$ and
\begin{equation}
  \label{eq:boundary_condition}
  X_0=2,\  X_1=0, \  X_2=1;\
   X_l=l, \  l\neq 0, 1, 2
\end{equation}
at $t\to +\infty$.  In the WKB approximation, the coupling constant
$\tilde J$ is given by
\begin{equation}
  \label{eq:exponential_accuracy}
  \tilde J = \tilde J^* \exp\left(-\frac{\tilde\eta}{\sqrt{na_B}}\right),
\end{equation}
where $\tilde\eta$ is the difference between the minimum action of the
cyclic exchange trajectory and the action of the trivial trajectory
$X_l(\tau)=l$.  We evaluate the instanton action and obtain
  $\tilde\eta$ numerically in Sec.~\ref{sec:numerics}.  The
  preexponential factor $\tilde J^*$ is discussed in
  Sec.~\ref{sec:preexp-fact}.

\subsection{Numerical evaluation of the instanton action}
\label{sec:numerics}

The full consideration of any exchange of particles in the Wigner
crystal requires that the motion of all particles forming the crystal
is taken into account. In practice, the consideration of a large
number of them is feasible. Assuming that in addition to $n$ particles
exchanging positions, $N$ spectators ($N/2$ on each side of the
exchanging cluster) are free to move from their equilibrium positions,
the numerical minimization of the dimensionless action
Eq.~(\ref{eq:eta_definition}) over the particle trajectories
$X_l(\tau)$ is straightforward to set up. The case of a single
instanton where two adjacent particles exchange positions has been
examined in detail previously.\cite{klironomos_exchange_2005} In
general, the problem is equivalent to solving a system of $n+N$
second-order differential equations of motion with appropriate
boundary conditions reflecting the initial and final state of the
system.

We take the opportunity here to introduce a shorthand notation for
relevant exchanges, denoting them by the index at static equilibrium
of the particles involved. For example, $[01]$ stands for the
two-particle exchange involving electrons labeled by $l=0$ and
1. Multiple exchanges can be chained together as we will see below.


\begin{figure*}[t]
\centering
\includegraphics[width=7.0in]{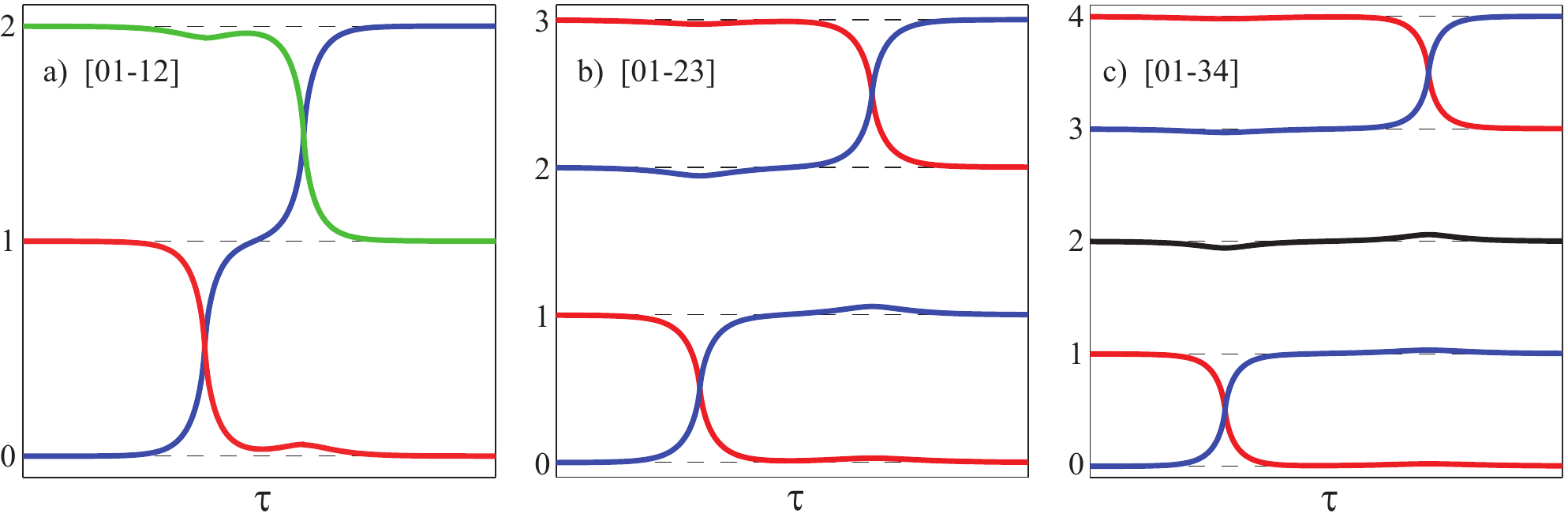}
\caption{\label{fig:exch_all}(Panel a): The exchange of three
particles as it unfolds in imaginary time. The numbers indicate the
equilibrium positions of the exchanging particles. We call this type
of exchange $[01\!-\!12]$ following the convention introduced in the
main text. The spectators, the other particles in the chain that
extend indefinitely above and below, are assumed frozen in place and
are not shown. (Panels b and c): The two most relevant exchanges in
addition to $[01\!-\!12]$. We dub them $[01\!-\!23]$ and
$[01\!-\!34]$, respectively, for obvious reasons. Note that in the
case of $[01\!-\!34]$, particle $2$ is, in fact, a spectator.}
\end{figure*}
One complication is the divergence of the bare Coulomb interaction
and the treatment of the associated singularity when two particles
occupy the same position in the course of the exchange. There are
various well-documented methods of how to treat such singularities
in order to obtain the solutions to the corresponding differential
equations of motion. It is simpler and better suited for our
purposes to consider a regularized Coulomb interaction by
introducing a small cutoff $\delta$:
\begin{equation}
\frac{1}{|X_l-X_{l'}|}\rightarrow\frac{1}{|X_l-X_{l'}|+\delta},
\end{equation}
and study the evolution of the solutions as $\delta\rightarrow 0$.

Considering two interacting instantons brings additional
complications in the calculation which will be explained below.
Starting with the simplest nontrivial exchange beyond
nearest-neighbor, one has to consider the exchange of three
particles which is illustrated in panel (a) of
Fig.~\ref{fig:exch_all}. Following the convention adopted above we
call it $[01\!-\!12]$ for the sake of brevity.

At sufficiently large temporal distance $\tau_0$ between the two
interacting instantons, the calculation is equivalent to that of two
single instantons, resulting in an exponent two times that of the
single instanton: $\eta_\infty=2\eta$; see Appendix
\ref{sec:analytical-treatment} for a more detailed exposition. At this
stage, the minimization of the dimensionless action
Eq.~(\ref{eq:eta_definition}), i.e. the solution of the differential
equations of motion, is carried out at fixed distance $\tau_0$, by
splitting the imaginary time interval into appropriate pieces and
joining the solutions using standard methods.

By examining the action as $\tau_0$ is varied, we find that the
instantons attract at large distances and repel at short distances.
The system permits an intuitive electrostatic analogy in terms of
interacting dipoles, which is further developed in Appendix
\ref{sec:analytical-treatment} [see
Eq.~(\ref{eq:instanton_interaction_electrostatic})]. Bringing the
instantons closer together, we discover that their interaction has a
minimum at a temporal distance $\tau_0$ of the order of the temporal
extent of the instanton. That characteristic distance $\tau_0$ depends
on the number of spectators that are allowed to move in addition to
the exchanging triad. As a result, the distance between the instantons
becomes an additional minimization parameter for the numerical
treatment of this problem. Therefore, the full calculation consists of
a minimization with respect to $\tau_0$ on top of each minimization of
the dimensionless action Eq.~(\ref{eq:eta_definition}).

Similar considerations apply for the more complicated exchanges
shown in panels (b) and (c) of Fig.~\ref{fig:exch_all}. Among a
large set of possible candidates, these are the ones that are
intuitively most relevant. Detailed calculations confirm that they
are negligible compared to the contribution of $[01\!-\!12]$.

\begin{figure}[tbh]
\centering
\includegraphics[width=3.41in]{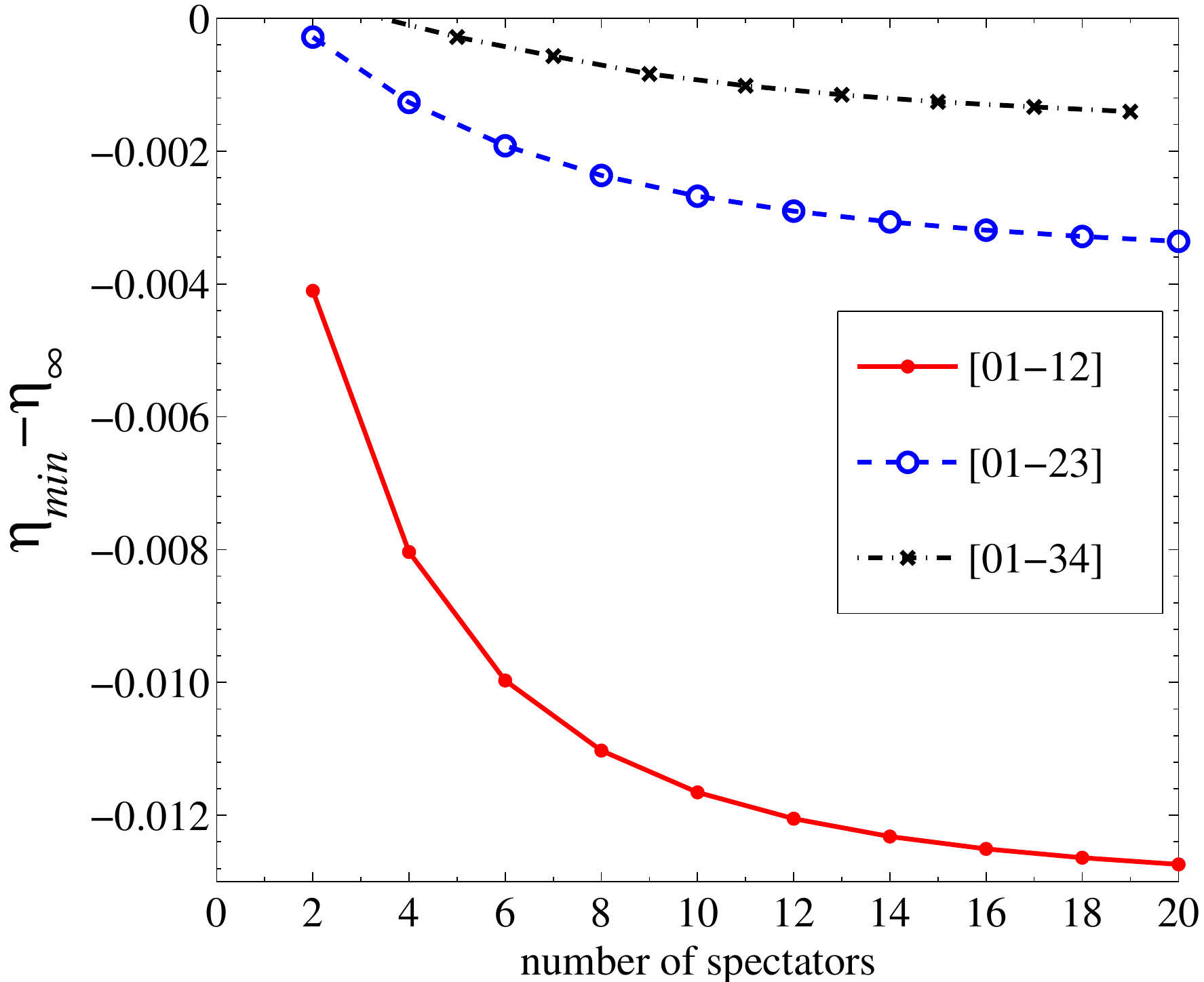}
\caption{\label{fig:comparison}A comparison of the difference
$\Delta\eta=\eta_{min}-\eta_\infty$ between the exponent obtained
for infinite distance $\tau_0$ between the instantons and that at
the minimum for the various kinds of exchanges considered here.
$[01\!-\!12]$, i.e. the three-particle exchange, dominates. The
curves shown have been obtained using a cutoff $\delta=10^{-3}$ for
the Coulomb interaction, which is more than adequate for this kind
of comparison. In fact, $[01\!-\!23]$ and $[01\!-\!34]$ have for all
practical purposes converged to their $\delta\rightarrow 0$ values.}
\end{figure}
Figure \ref{fig:comparison} shows how the difference
$\Delta\eta=\eta_{min}-\eta_\infty$ between the exponent corresponding
to infinite distance $\tau_0$ between the instantons and that obtained
at the minimum depends on the number of spectators and the cutoff used
in the calculations. A cutoff $\delta=10^{-3}$ is more than adequate for this
comparison, and it in fact gives an excellent approximation to the
result one would obtain with an unscreened Coulomb interaction for
$[01\!-\!23]$ and $[01\!-\!34]$. Given the considerable separation
between the curves, it is reasonable to claim that even at the
$\delta\rightarrow 0$ limit, the exchange $[01\!-\!12]$ of three
consecutive particles is the dominant one.

An important observation that we should make at this point is that
while the exponents themselves converge quite slowly to their
$\delta\rightarrow 0$ values, the difference
$\Delta\eta=\eta_{min}-\eta_\infty$ between the exponent obtained
for infinite distance $\tau_0$ between the instantons and that at
the minimum converges much faster. Typically, one would have to go
down to $\delta\sim10^{-7}$ to obtain a value of the exponent that
is in reasonable agreement with the result of the calculation using
an unscreened Coulomb interaction. The calculation with
$\delta\sim10^{-4}$ gives an excellent approximation to the result
for $\Delta\eta$ one would obtain with an unscreened Coulomb
interaction. Use of the regularized Coulomb interaction simplifies
the calculation significantly from two aspects: computational cost
and complexity of the code involved.

\begin{figure}[tbh]
\centering
\includegraphics[width=3.41in]{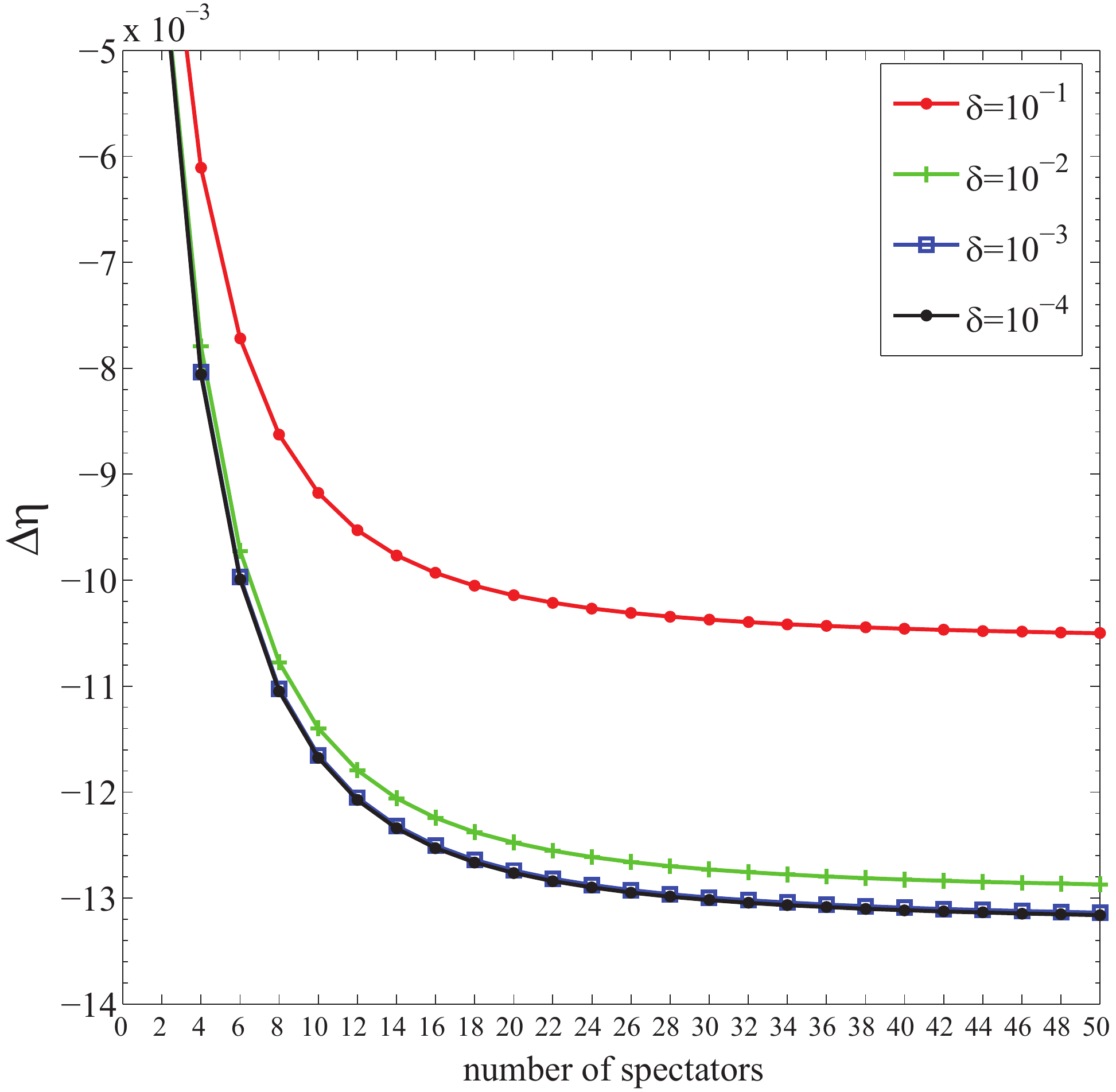}
\caption{\label{fig:deltaeta_vs_spec}The difference
$\Delta\eta=\eta_{min}-\eta_\infty$ for the dominant three-particle
exchange as a function of the cutoff $\delta$ used in the
calculation. The result for $\delta=10^{-4}$ is an excellent
approximation to the result one would obtain with an unscreened
Coulomb interaction.}
\end{figure}

Focusing on the three-particle exchange, we were able to extend the
calculation to $25$ spectators on each side of the exchanging triad,
$50$ spectators total. Figure \ref{fig:deltaeta_vs_spec} shows the
evolution of $\Delta\eta$ as a function of the cutoff $\delta$ and
the number of spectators that are allowed to move. We find that the
value quickly saturates for decreasing $\delta$. As mentioned above,
the calculation with $\delta\sim10^{-4}$ gives an excellent
approximation to the result one would obtain with an unscreened
Coulomb interaction.

With this extended data set, it is possible to extract the
asymptotic behavior in the limit of large $N$. Neglecting
logarithmic corrections, an excellent approximation is the
expression taken from Ref.~[\onlinecite{klironomos_exchange_2005}]:
\begin{equation}
\label{eq:asympt} (\Delta\eta)_N=\Delta\eta+\frac{\alpha}{N^2}.
\end{equation}
The above formula results in excellent fits, and we obtain
$\Delta\eta_{0112}=-0.01324\pm0.00001$, and
$\alpha_{0112}=0.21\pm0.01$, in the limit where all particles
participate in the three-particle exchange.

Using a similar procedure and by comparing the value obtained from
the extended data set and that shown in Fig.~\ref{fig:comparison} we
obtain estimates for the parameters relevant to the other two
exchanges, $[01\!-\!23]$ and $[01\!-\!34]$. In particular, we find
that $\Delta\eta_{0123}=-0.0036\pm0.0002$,
$\alpha_{0123}=0.11\pm0.05$ and that
$\Delta\eta_{0134}=-0.0016\pm0.0002$, $\alpha_{0134}=0.08\pm0.05$.

\subsection{Pre-exponential factor}
\label{sec:preexp-fact}

As we saw in Sec.~\ref{sec:numerics}, the dimensionless instanton
action $\tilde\eta<2\eta$.  This fact is important for the theory of
equilibration of the Wigner crystal, see
Sec.~\ref{sec:rate-full-equil}.  In terms of the instanton
trajectories, it is a consequence of the attraction of single
instantons at large temporal distances.  This attraction can be
understood analytically, see Appendix~\ref{sec:analytical-treatment}.
On the other hand, the attraction is very weak, $\Delta\eta\sim
10^{-2}$.  This implies that the three particle instanton consists of
two single instantons, which are not significantly distorted by the
interaction with each other, see Fig.~\ref{fig:exch_all}(a).  The
approximation of two weakly coupled instantons enables us to find the
pre-exponential factor $\tilde J^*$ in the expression
(\ref{eq:exponential_accuracy}) for the exchange constant $\tilde J$
while avoiding the evaluation of the fluctuation determinant near the
instanton trajectory.

We start by considering the contribution of a single instanton to the
path integral representing the evolution operator
\begin{equation}
  \label{eq:single-instanton_integral}
  I\int_0^{\mathcal T}dt=\frac{J}{2\hbar}\mathcal T.
\end{equation}
Here $I$ includes the instanton action and the integral over the
Gaussian fluctuations near the instanton trajectory; the integral
over the position of the instanton $\tau$ accounts for the zero mode.
The right-hand side of Eq.~(\ref{eq:single-instanton_integral}) is
obtained by isolating the contribution of the nearest neighbor
exchange in the Hamiltonian (\ref{eq:H_approximated}) and expanding
the evolution operator $e^{-H\mathcal{T}/\hbar}$ to first order in
$J$.  Equation (\ref{eq:single-instanton_integral}) enables us to
identify $I=J/2\hbar$.

\begin{figure}[tbh]
\centering
\includegraphics[width=3.41in]{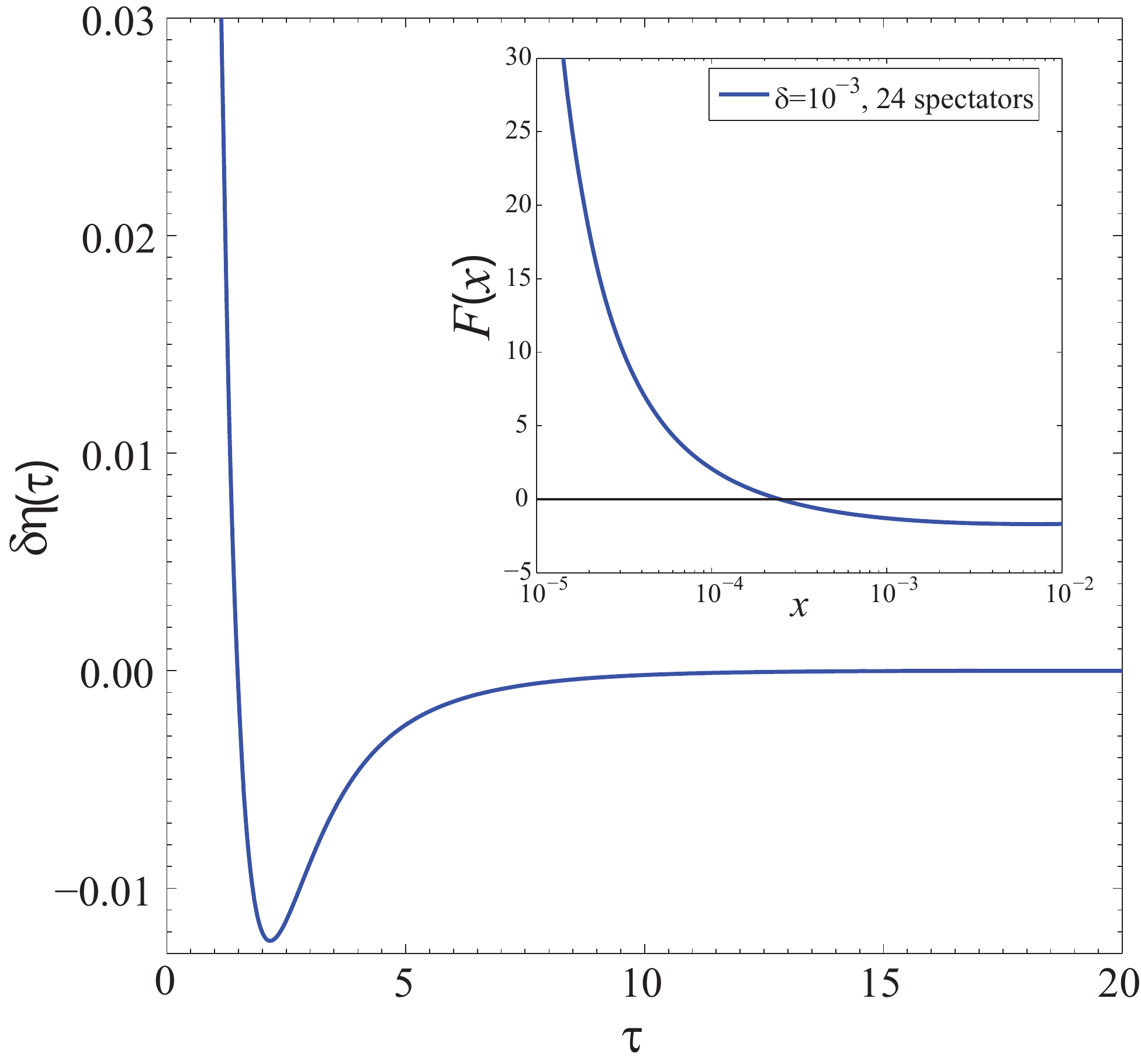}
\caption{\label{fig:prefactor_all} Interaction between two instantons
$\delta \eta (\tau)= \eta(\tau) - \eta_\infty$ at a temporal distance $\tau$ evaluated numerically.
The inset shows the numerically evaluated function $F(x)$ defined by Eq.~(\ref{eq:F}).}
\end{figure}

Next, we consider the contribution to the path integral due to a
single three-particle exchange shown in Fig.~\ref{fig:exch_all}(a)
\begin{equation}
  \label{eq:double-instanton_integral}
  I^2\!\int_0^{\mathcal T}\!dt_1\int_{\tau_1}^{\mathcal T}\!dt_2\,
     e^{-\delta\eta(\tau_2-\tau_1)/\sqrt{na_B}}
  =\frac{\tilde J\mathcal T}{2\hbar}
   +\frac12\bigg(\frac{J\mathcal T}{2\hbar}\bigg)^2.
\end{equation}
Here the relation between $t_{1,2}$ and $\tau_{1,2}$ is given by
Eq.~(\ref{eq:rescaled_variables}). The quantity $\delta \eta(\tau)$ is
defined as the difference between the action of a two-instanton
trajectory with the distance $\tau$ between the instantons and the
action $\eta_\infty=2 \eta$ of two instantons at infinite separation,
$\tau\to \infty$; its minimum value coincides with $\Delta\eta$. The
numerically evaluated $\delta \eta(\tau)$ is shown in
Fig.~\ref{fig:prefactor_all}.  In the left-hand side of
Eq.~(\ref{eq:double-instanton_integral}) we neglected the effect of
the interaction between the instantons on the Gaussian fluctuations
about the instanton trajectories.  In the right-hand side we accounted
for the fact that three-particle exchange processes appear as a result
of the three-particle permutation operators
(\ref{eq:three-particle_permutation}) in the Hamiltonian, or in the
second order perturbation theory in the nearest-neighbor exchange $J$.
Using the relation $I=J/2\hbar$ we express $\tilde J$ as
\begin{equation}
  \label{eq:tilde_J_integral}
  \tilde J = J^2 \frac{\varepsilon a_B}{2(na_B)^{3/2}e^2} F(na_B),
\end{equation}
where
\begin{equation}
  \label{eq:F}
  F(x)=\int_0^{\infty}
       \left[\exp\left(-\frac{\delta\eta(\tau)}{\sqrt{x}}\right)-1\right]
       d\tau.
\end{equation}
The integral (\ref{eq:F}) converges because, as discussed in
Appendix~\ref{sec:analytical-treatment}, at large $\tau$ the
interaction of single instantons falls off as $\delta\eta(\tau)\propto
\tau^{-2}$, see Eq.~(\ref{eq:instanton_interaction_electrostatic}).

At $x\to0$ the integral (\ref{eq:F}) can be evaluated in the saddle
point approximation
\begin{equation}
  \label{eq:saddle-point}
  F(x)=\sqrt{\frac{2\pi}{\delta\eta''}}\,x^{1/4} e^{-\Delta\eta/\sqrt{x}},
\end{equation}
where $\Delta\eta\approx-0.01324$ and $\delta\eta'' \approx0.024$ are
the values of $\delta\eta(\tau)$ and its second derivative at the
minimum.  Substituting Eq.~(\ref{eq:saddle-point}) into
(\ref{eq:tilde_J_integral}) we find that $\tilde J$ depends on $na_B$
as
\begin{equation}
  \label{eq:tildeJ_large_rs}
  \tilde J \sim (na_B)^{5/4}\frac{e^2}{\varepsilon a_B}
    \exp\left(-\frac{2\eta+\Delta\eta}{\sqrt{na_B}}\right).
\end{equation}
Note that the sign of the exchange constant is positive, and the
pre-exponential factor up to a numerical factor coincides with the one
for the nearest neighbor exchange (\ref{eq:J}).  This is the expected
behavior for any exchange process controlled by a well-defined
instanton trajectory.

It is important to note that the saddle point approximation (\ref{eq:saddle-point}) holds only
at $x\ll(\Delta\eta)^2$.  In physical terms this means $na_B\ll
10^{-4}$.  On the other hand, the WKB approximation is applicable
under the less stringent condition $na_B\ll1$.  To evaluate the integral
(\ref{eq:F}) in the intermediate region $10^{-4}\ll x\ll 1$ one can
completely neglect the attraction of single instantons and the shallow
minimum of $\delta\eta(\tau)$ associated with it.  Instead, one should
notice the strong repulsion of the cores of the instantons at short
distances.  Taking this repulsion into consideration, one can
approximate the integrand of Eq.~(\ref{eq:F}) by
$-\theta(\tau_0-\tau)$, where $\theta(x)$ is the unit step function,
and $\tau_0(x)$ is defined as the solution of the equation
$\delta\eta(\tau_0)=x^{1/2}$.  This yields
\begin{equation}
  \label{eq:F_via_tau0}
  F(x)=-\tau_0(x).
\end{equation}
The cores of single instantons repel exponentially,
$\delta\eta(\tau)\sim e^{-\lambda\tau}$ with $\lambda\sim 1$.  In that
case $\tau_0(x)$ depends on $x$ logarithmically, and we find
\begin{equation}
  \label{eq:F_with_log}
  F(x)=-\frac{1}{2\lambda}\ln\frac{1}{x}.
\end{equation}
Interestingly, the results (\ref{eq:F_via_tau0}) and
(\ref{eq:F_with_log}) are negative.  This means that the next nearest
neighbor exchange constant (\ref{eq:tilde_J_integral}) has
ferromagnetic sign only at $na_B<x^*$, where $x^*\sim10^{-4}$, and becomes
antiferromagnetic at higher densities, $na_B>x^*$.  In this regime
\begin{equation}
  \label{eq:tildeJ_moderate_rs}
  \tilde J\sim -J^2 \frac{\varepsilon a_B}{(na_B)^{3/2}e^2}\ln\frac{1}{na_B}.
\end{equation}
The result of numerical evaluation of $F(x)$ is shown in Fig.~\ref{fig:prefactor_all}.
The sign of the exchange constant (\ref{eq:tilde_J_integral}) changes at $na_B=x^*\approx 3 \times 10^{-4}$.

\section{Rate of full equilibration of the Wigner crystal at low temperatures}
\label{sec:rate-full-equil}

In Secs.~\ref{sec:corrections} and \ref{sec:WKB} we studied the
corrections to the leading-order Hamiltonian of the Wigner crystal
given by Eqs.~(\ref{eq:H_rho}) and (\ref{eq:spin-chain}).  These
corrections break the integrability of the problem and give rise to the
scattering of excitations.  As a result, the system relaxes to
thermodynamic equilibrium.  In this section we study the corresponding
equilibration rate in the regime of low temperatures, $T\ll J$.

This problem has been solved previously for weakly interacting
electrons,\cite{micklitz_transport_2010} for the spinless Wigner
crystal,\cite{matveev_equilibration_2010} as well as for a spinless
quantum liquid with arbitrary interaction
strength.\cite{matveev_equilibration_2012, matveev_scattering_2012,
  matveev_equilibration_2013} The equilibration proceeds in two
stages.  First, the low-energy excitations collide with each other and
achieve thermal equilibrium.  At low temperatures the momenta of these
excitations are small, $p\ll \hbar n$, and to a first approximation the
collisions conserve the total momentum of the excitations $P_{\rm
  ex}$.  As a result, the equilibrium state of the gas of excitations
is characterized by $P_{\rm ex}\neq0$.  This first stage of the
equilibration process proceeds relatively quickly, with the typical
relaxation time $\tau_0$ following a power-law temperature dependence.

The second stage of the equilibration process involves slow relaxation
of $P_{\rm ex}$ to its equilibrium value.  For a system at rest, the
total momentum of the excitations in equilibrium is zero, and the
approach to equilibrium follows the usual relaxation law
\begin{equation}
  \label{eq:momentum_relaxation}
  \dot P_{\rm ex}=-\frac{P_{\rm ex}}{\tau}.
\end{equation}
The microscopic processes driving the relaxation
(\ref{eq:momentum_relaxation}) involve excitations diffusing in
momentum space from the vicinity of one Fermi point to the other.  The
bottleneck in this process is the center of the Fermi sea where the
excitation energy reaches its maximum value $\Delta$.  As a result,
the rate of full equilibration follows the activated temperature
dependence $\tau^{-1}\propto e^{-\Delta/T}$.

In a spinless system, the excitation with the lowest energy in the
center of the Fermi sea is essentially a hole dressed by
electron-electron interactions.  The full expression for the
equilibration rate is given by\cite{matveev_equilibration_2012}
\begin{equation}
  \label{eq:equilibration_rate_spinless}
  \tau^{-1}= \frac{3\hbar k_F^2  B}{\pi^2\sqrt{2\pi m^*T}}
            \left(\frac{\hbar v}{T}\right)^3  e^{-\Delta/T}.
\end{equation}
Here, $k_F$ is the Fermi wave vector and $v$ is the velocity of the
low-energy excitations.  The parameters $m^*$ and $B$ are,
respectively, the effective mass and the diffusion constant in
momentum space for the hole excitation at the center of the Fermi sea.
The temperature dependence of the diffusion constant is given by
\begin{equation}
  \label{eq:chi_spinless}
  B=\frac{4\pi^3 n^2 T^5}{15\hbar^5m^2v^8}
       \left(
        \Delta''-\frac{2v'}{v}\Delta'+\frac{\Delta'^2}{m^*v^2}
       \right)^2.
\end{equation}
The primes in Eq.~(\ref{eq:chi_spinless}) denote derivatives with
respect to the particle density $n$.  We shall now discuss how these
results change in the presence of spins.

\subsection{Equilibration rate for a Wigner crystal at $T\ll J$}
\label{sec:equil-rate-wign}

The general picture of two-stage relaxation also applies to systems
with spins.  In this case there are two types of low-energy
excitations in the system, corresponding to the charge and spin
degrees of freedom.  For instance, at strong interactions the charge
excitations are the phonons in the Wigner crystal,
cf.~Eq.~(\ref{eq:H_rho_phonons}), whereas the elementary excitations
of the spin Hamiltonian (\ref{eq:spin-chain}) are the so-called
spinons with the excitation
spectrum\cite{des_cloizeaux_spin-wave_1962, faddeev_what_1981}
\begin{equation}
  \label{eq:spinon_spectrum}
  \epsilon(q)=\frac{\pi J}{2}\sin q.
\end{equation}
Here the wavevector of the excitations on the spin chain is related to
their physical momentum as $q=p/\hbar n$.

\begin{figure}[tbh]
\centering
\includegraphics[width=0.45\textwidth]{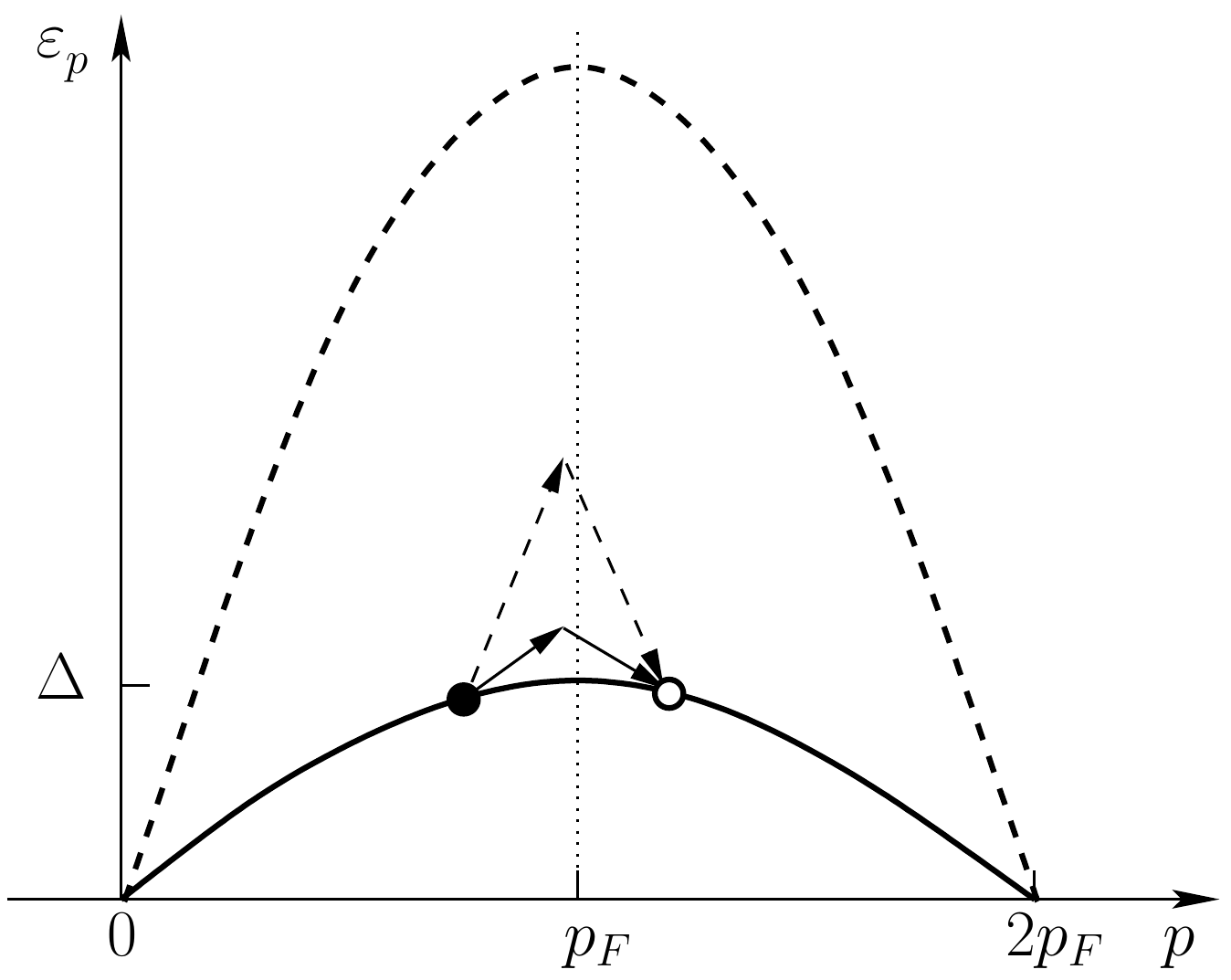}
\caption{\label{fig:hole-parabola7} The spectrum of elementary
  excitations of a one-dimensional Wigner crystal consists of two
  branches, phonons and spinons, shown by dashed and solid lines,
  respectively.  Equilibration of the system is controlled by the
  scattering processes in which a quasiparticle crosses the edge of
  the Brillouin zone $p=p_F$, i.e., umklapp processes.  At low
  temperatures the occupation numbers of the excitations at the edge
  of the Brillouin zone are exponentially small.  Because the energy
  of a spinon with momentum $p_F$ is much smaller than that of a
  phonon, the contribution of the latter is negligible.  Each umklapp
  process includes absorption of one acoustic excitation and emission
  of another one.  Processes involving two acoustic phonons and two
  acoustic spin excitations are shown by thin dashed and solid lines,
  respectively.  Mixed processes involving one of each type of
  acoustic excitations are also allowed.}
\end{figure}

Because of the smallness of the exchange constant $J$ in
Eq.~(\ref{eq:spinon_spectrum}), the spinons are the lowest-energy
excitations of the system at any given momentum, see
Fig.~\ref{fig:hole-parabola7}.  Thus the full equilibration of the
Wigner crystal is achieved by their diffusion in momentum space,
similar to that of the holes in spinless systems.  We therefore
conclude that the activation energy $\Delta$ that will appear in the
generalization of Eq.~(\ref{eq:equilibration_rate_spinless}) to the
spin-degenerate case will be given by the maximum energy of the
spinon, and the effective mass will be determined by the curvature of
$\epsilon(q)$ near the maximum,
\begin{equation}
  \label{eq:Delta_and_mass}
  \Delta=\frac{\pi J}{2},
\quad
  \frac{1}{m^*}=\frac{\pi J}{2\hbar^2 n^2}.
\end{equation}

At $q$ near 0 or $\pi$ the spinon spectrum is linear with the velocity
\begin{equation}
  \label{eq:v_sigma}
  v_\sigma=\frac{\pi J}{2\hbar n}.
\end{equation}
Such low-energy spin
excitations can be equivalently represented in terms of the bosons in
the Tomonaga-Luttinger liquid, which for particles with spin is
described by the Hamiltonian\cite{giamarchi2004quantum}
\begin{equation}
  \label{eq:Luttinger_with_spin}
  H_{\rm TL} = \sum_p \left[v_\rho|p|b_p^\dagger b_p^{}
                       +v_\sigma|p|c_p^\dagger c_p^{}\right].
\end{equation}
Here $b_p$ and $c_p$ are the bosonic annihilation operators of the
charge and spin excitations, while $v_\rho$ and $v_\sigma$ are the
corresponding velocities.

At the first stage of the equilibration process the bosonic
excitations collide with each other, and their distribution function
relaxes to the equilibrium form
\begin{equation}
  \label{eq:Bose_distribution}
  N_p^{(\rho,\sigma)}=\frac{1}{e^{(v_{\rho,\sigma}|p|-up)/T}-1}.
\end{equation}
These equlibration processes are caused by the integrability-breaking
perturbations (\ref{eq:perturbation}),
(\ref{eq:three-particle_perturbation}), and (\ref{eq:v_rho_sigma_1}).
Because typical excitations participating in this process have
energies of order temperature, the relaxation rate $\tau_0^{-1}$ is
only power-law small in $T$.  Compared with the exponentially slow
rate of full relaxation (\ref{eq:equilibration_rate_spinless}), these
processes can be considered instantaneous, and the magnitude of
$\tau_0^{-1}$ has no effect on the subsequent discussion.

The parameter $u$ in Eq.~(\ref{eq:Bose_distribution}) accounts for
momentum conservation in the boson collisions.  The total momentum of
the gas of excitations is then easily found,
\begin{equation}
  \label{eq:momentum_of_excitations}
  P_{\rm ex}=\frac{\pi L T^2}{3\hbar}
           \left(\frac{1}{v_\rho^3}+\frac{1}{v_\sigma^3}\right)u.
\end{equation}
It is important to keep in mind that
Eq.~(\ref{eq:momentum_of_excitations}) is not the full momentum of the
system.  In the Luttinger liquid theory the latter is given
by\cite{haldane_luttinger_1981}
\begin{equation}
  \label{eq:full_momentum}
  P=p_F(N^R-N^L)+P_{\rm ex},
\end{equation}
where the zero modes $N^R$ and $N^L$ have the meanings of the total
numbers of the right- and left-moving electrons in the system.  The
first term in Eq.~(\ref{eq:full_momentum}) accounts for the simple
fact that even in the absence of excitations the system can move as a
whole and thus have a nonvanishing momentum.

At $u>0$ the occupation numbers of the bosonic acoustic excitations
(\ref{eq:Luttinger_with_spin}) depend on the direction of motion, with
the right-moving states being more populated than the left-moving ones
at the same energy.  Thus when spinons with momenta near $p_F$ collide
with acoustic excitations, their momentum is more likely to increase
than decrease.  This gives rise to a net current of spinons in
momentum space through the edge of the Brillouin zone $p=p_F$, see
Fig.~\ref{fig:hole-parabola7}.  As a result of such umklapp scattering
processes the right-moving spinons convert to left-moving ones,
leading to a decrease of velocity $u$.  Since the total momentum
(\ref{eq:full_momentum}) of the system is conserved, during this
second stage of the equilibration process the momentum is being
transferred from excitations to the zero
modes.\cite{matveev_equilibration_2011, matveev_equilibration_2013}
This corresponds to backscattering of electrons in the system.  The
resulting effects on the electronic transport are discussed in
Sec.~\ref{sec:summary}.

At low temperature $T\ll\Delta$ the typical change of momentum
$T/v_{\rho,\sigma}$ in the processes shown in
Fig.~\ref{fig:hole-parabola7} is small compared to the typical scale
$\sqrt{m^*T}$ at which the spinon distribution function varies near
the edge of the Brillouin zone.  This enables one to evaluate the rate
of change of the momentum of excitations $\dot P_{\rm ex}$ using the
Fokker-Planck equation for the spinon distribution function
\begin{equation}
  \label{eq:Fokker-Planck}
  \partial_t f = -\partial_p J,
\quad
  J=-\frac{B}{2}\left[\frac{\varepsilon_p'}{T}+\partial_p\right]f,
\end{equation}
where $\varepsilon_p=\epsilon(p/\hbar n)$.  By imposing the boundary
conditions on the distribution function obtained with the help of
Eq.~(\ref{eq:Bose_distribution}), one finds
\begin{equation}
  \label{eq:Pdot}
  \dot P_{\rm ex}=-u
    \frac{2L\hbar^3 k_F^2 B}{\pi T\sqrt{2\pi m^* T}}\,
    e^{-\Delta/T}.
\end{equation}
The derivation is identical to the case of diffusion of holes in a
spinless system,\cite{matveev_equilibration_2010,
  matveev_equilibration_2012, matveev_equilibration_2013} with the
exception of a factor of 2 accounting for two possible spin
polarizations of spinons.  In this approach the diffusion constant $B$
appears as a phenomenological parameter.  Its evaluation requires a
microscopic treatment of the coupling of the spinons to the bosonic
excitations, and will be discussed below.

We can now combine the results (\ref{eq:momentum_of_excitations}) and
(\ref{eq:Pdot}) with the definition (\ref{eq:momentum_relaxation}) of
the equilibration time $\tau$.  The resulting expression coincides
with Eq.~(\ref{eq:equilibration_rate_spinless}) with the substitution
$v^3\to 2/(v_\rho^{-3}+v_\sigma^{-3})$.  Taking advantage of the fact
that in the Wigner crystal $v_\sigma\ll v_\rho$ and using
Eqs.~(\ref{eq:Delta_and_mass}) and (\ref{eq:v_sigma}), the result can
be brought to the form
\begin{equation}
  \label{eq:equilibration_rate_with_spin}
  \tau^{-1}=\frac{3\pi^3B}{32 n^2}
                 \left(
                    \frac{J}{T}
                 \right)^{7/2}
                 \exp
                 \left(
                   - \frac{\pi J}{2T}
                 \right).
\end{equation}
In a system described by the leading-order Hamiltonian (\ref{eq:H_rho}),
(\ref{eq:spin-chain}) integrability results in a vanishing
diffusion constant $B$.  To evaluate it, one has to consider the
corrections to the Hamiltonian discussed in Sec.~\ref{sec:corrections}.

\subsection{Diffusion of spinons in momentum space}
\label{sec:diff-spin-moment}

The parameter $B$ in the expression
(\ref{eq:equilibration_rate_with_spin}) for the equilibration rate has
the meaning of the diffusion constant describing the motion of spinons
in momentum space near the top of the spinon spectrum.  It is formally
defined\cite{matveev_equilibration_2012} as
\begin{equation}
  \label{eq:B_definition}
  B=\frac{1}{\hbar^2}\sum_{\delta p} (\delta p)^2 W(\delta p),
\end{equation}
where $W(\delta p)$ is the rate of the scattering events changing the
spinon momentum by $\delta p$.  The scattering originates from the
interaction of the spinon with the low-energy excitations of the
system described by Eq.~(\ref{eq:Luttinger_with_spin}).

It is convenient to classify the low-energy excitations as belonging
to one of the four branches, $\rho R$, $\rho L$, $\sigma R$, and
$\sigma L$, depending on their charge or spin nature, and the direction
of motion.  Because the velocity of the spinon near the top of the
spectrum is small compared to both $v_\sigma$ and $v_\rho$, scattering
processes involving excitations in only one of the four branches are
forbidden by conservation of momentum and energy.  Thus, the dominant
scattering processes involve two branches, and the scattering rate has
the following general form
\begin{equation}
  \label{eq:W}
  W(\delta p)=\frac{1}{2}\sum_{\alpha,\beta}\sum_{a,e} W_{ae}^{\alpha\beta}(\delta p).
\end{equation}
Here $a$ and $e$ label the branches from which excitations are
absorbed and emitted, respectively, whereas $\alpha$ and $\beta$
denote the spin projections of the spinon before and after the
scattering event.  The partial scattering rates are obtained from
Fermi's golden rule,
\begin{equation}
  \label{eq:W^ae}
  W_{ae}^{\alpha\beta}(\delta p)=\frac{2\pi}{\hbar}\sum_{p_a,p_e}\!
                  M^{\alpha\beta}_{p_ap_e}(\delta p)\,
                  \delta(\epsilon_p-\epsilon_{p+\delta p}+v_ap_a-v_ep_e).
\end{equation}
Here $v_a$ and $v_e$ denote the velocities of the absorption and
emission branches, which according to
Eq.~(\ref{eq:Luttinger_with_spin}) are $v_{\rho R}=-v_{\rho L}=v_\rho$
and $v_{\sigma R}=-v_{\sigma L}=v_\sigma$.  Momentum conservation is
ensured by the matrix element of the $T$-matrix entering the
definition
\begin{eqnarray}
  \label{eq:M_definition}
  M^{\alpha\beta}_{p_ap_e}(\delta p)&=&\sum_{i,f}w_i
   |\langle f;\beta,p+\delta p|\hat T|i;\alpha,p\rangle|^2
\nonumber\\
  &&\times
  \delta_{P_a^f,P_a^i-p_a}  \delta_{P_e^f,P_e^i+p_e}.
\end{eqnarray}
In this expression $i$ and $f$ refer to the initial and final states
of the Luttinger liquid, $w_i$ is the Gibbs weight of the initial
state, $p$ and $p+\delta p$ are the values of the spinon momentum
before and after the collision. Finally, $P_a^{i(f)}$ is the total
initial (final) momentum of the excitation branch $a$, and
$P_e^{i(f)}$ is that for branch $e$.

Each of the indices $a$ and $e$ in Eq.~(\ref{eq:W}) takes one of the
four values, $\rho R$, $\rho L$, $\sigma R$, and $\sigma L$, resulting
in 16 possible contributions to the scattering rate.  It is convenient
to group these contributions into three classes, determined by the
charge or spin nature of the two branches participating in spinon
scattering.  Accordingly, the diffusion constant
(\ref{eq:B_definition}) is presented as a sum of three contributions
\begin{equation}
  \label{eq:B_three_channels}
  B=B_{\rho\rho}+B_{\rho\sigma}+B_{\sigma\sigma}.
\end{equation}
Here $B_{\rho\rho}$ accounts for the four types of processes in which
$a$ and $e$ are chosen from the branches $\rho R$ and $\rho L$, the
contribution $B_{\sigma\sigma}$ accounts for the four terms in which
only $\sigma R$ and $\sigma L$ branches are involved, and
$B_{\rho\sigma}$ includes the eight remaining types of processes.
To evaluate the three contributions to $B$ we need to consider the coupling of the spinon to the acoustic spin and charge excitations.

\subsubsection{Effective Hamiltonian of the spinon interacting with acoustic excitations}
\label{sec:effective_Hamiltonian}

The general form of the Hamiltonian of the Wigner crystal discussed
Sec.~\ref{sec:corrections} is valid in a rather wide temperature
interval $T\ll(e^2/\varepsilon a_B)(na_B)^{7/4}$. In order to discuss
equilibration of the system at $T \ll J$ where the density of
high-energy spinons is exponentially small it is sufficient to
consider the interaction of a single spinon with acoustic spin and
charge excitations. To this end we introduce an effective Hamiltonian
describing the motion of a spinon coupled to the acoustic modes.

In the low-energy limit the acoustic modes are described by the Tomonaga-Luttinger Hamiltonian (\ref{eq:Luttinger_with_spin}).
In the effective theory, the spinon is treated as a mobile impurity with  spectrum (\ref{eq:spinon_spectrum}) and spin $\bm S$.
The Hamiltonian of a free spinon can be written as
\begin{equation}\label{eq:H_0_spinon}
    H^{(0)}_{\mathrm{sp}}=\epsilon\left( -i a\partial_Y \right)\simeq \Delta -\frac{\hbar^2\left( -i \partial_Y -\pi n/2\right)^2}{2m^*}.
\end{equation}
The last expression in Eq.~(\ref{eq:H_0_spinon}) is valid near the top of the spinon spectrum, and the values of $\Delta$ and $m^*$ are given by Eq.~(\ref{eq:Delta_and_mass}).

The parameter $\Delta=\pi J/2$ in Eq.~(\ref{eq:H_0_spinon}) depends on the particle density $n$. This enables us to obtain the coupling of the spinon to the phonons following the procedure of Sec.~\ref{sec:coupling-charge-spin},
\begin{equation}\label{eq:spinon_phonon_coupling}
    \hat V_{\mathrm{ph}}=- n u' (Y)\Delta',
\end{equation}
where  $\Delta'=d\Delta/dn$, cf. Eq.~(\ref{eq:chi_spinless}).

Quite generally, the coupling of an impurity with spin $\bm S$ to the
low-energy spin degrees of freedom may be described by a perturbation
of the form
\[
  \hat V_K=J_K^R\,\bm S\cdot \bm s^R(Y)
           +J_K^L\,\bm S\cdot \bm s^L(Y).
\]
Here $Y$ is the position of the spinon in Lagrangian variables defined
as $y=l a$.\cite{matveev_scattering_2012} A Hamiltonian of this form
was applied recently to the related problem of a mobile impurity with
spin in a one-dimensional Fermi gas.\cite{lamacraft_kondo_2008} The
spin densities $\bm s^R(Y)$ and $\bm s^L(Y)$ associated with the
right- and left-moving excitations at the position $Y$ of the impurity
are easily expressed in terms of the creation and annihilation
operators of the one-dimensional fermions considered in
Ref.~\onlinecite{lamacraft_kondo_2008}.  Upon bosonization, the
expressions for the $z$ components take the forms
\begin{equation}
  \label{eq:s_z}
  s_z^{R,L}(Y) =\mp \frac{i}{2}\sum_p\sqrt{\frac{|p|}{\pi\hbar L}}\,
            \theta(\pm p)
            \big(c_p e^{ipY/\hbar}-c_p^\dagger e^{-ipY/\hbar}\big).
\end{equation}
By virtue of the universality of the Luttinger liquid theory, this
expression applies at any strength of interaction between electrons.
The expressions for the $x$ and $y$ components of the spin densities
are more complicated, but their explicit forms will not be used in
this paper. For the spinon near the top of the spectrum, the constants $J_K^R$
and $J_K^L$ are equal to each other and will be denoted by $J_K$. Thus, the operator $\hat V_K$ takes the form
\begin{equation}\label{eq:Kondo_top}
      \hat V_K=J_K\,\bm S\cdot \left[\bm s^R(Y)
           + \bm s^L(Y)\right].
\end{equation}

The operators (\ref{eq:spinon_phonon_coupling}) and (\ref{eq:Kondo_top}) describe the coupling of the spinon to a single bosonic excitation. In principle, the spinon can interact with an arbitrary number of bosons. Of all such perturbations we will only need the explicit form of the operator that couples the spinon to one charge- and one spin-excitation.
It can be obtained by noticing that the coupling constant $J_K$ in Eq.~(\ref{eq:Kondo_top}) depends on density and applying the procedure of Sec.~\ref{sec:coupling-charge-spin},
\begin{equation}\label{eq:J_Kondo_prime}
    \hat V_2=- n u' (Y)J'_K\,\bm S\cdot \left[\bm s^R(Y)
           + \bm s^L(Y)\right].
\end{equation}

Finally, the evaluation of the diffusion constant $B$ requires consideration of the coupling of the acoustic spin and charge modes  to each other.
We start by rewriting
the spin part of the Hamiltonian (\ref{eq:Luttinger_with_spin}) in terms of the $z$-components of spin density operators (\ref{eq:s_z}) as
\begin{equation}\label{eq:TL_spin}
    \sum_p v_\sigma|p|c_p^\dagger c_p^{}= 2\pi \hbar \int v_\sigma \left\{ \left[s_z^L(y) \right]^2 + \left[s_z^R(y) \right]^2\right\}d y .
\end{equation}
Noticing that $v_\sigma$ depends on the electron density and applying the procedure of Sec.~\ref{sec:coupling-charge-spin} again we obtain the perturbation of the form
\begin{equation}\label{eq:V_rho_sigma}
    \hat V_{\rho\sigma}= - 2\pi \hbar n v'_\sigma \int u'(y)\left\{ \left[s_z^L(y) \right]^2 + \left[s_z^R(y) \right]^2\right\}d y.
\end{equation}
It is important to note that the operator (\ref{eq:V_rho_sigma})
couples a phonon to two spin excitations moving in the same
direction. The Hamiltonian also contains the perturbation that couples
a single phonon to one right-moving and one left-moving spin boson. To
obtain its explicit form we need to consider the
correction\cite{giamarchi2004quantum} to the spin sector of the
Tomonaga-Luttinger Hamiltonian (\ref{eq:TL_spin}),
\begin{equation}\label{eq:H_g}
    H_g= -2\pi \hbar g v_\sigma \int {\bm s}^R(y) \cdot{\bm s}^L(y)   dy  .
\end{equation}
At low energies this perturbation scales to zero logarithmically, $g=1/\ln (J/T)$. The density dependence of $v_\sigma$ yields a coupling of the form
\begin{equation}\label{eq:V_g}
    \hat{V}_g=2\pi \hbar g n v_\sigma' \int u'(y)\,{\bm s}^R(y) \cdot{\bm s}^L(y)   dy  .
\end{equation}

To summarize, in the effective low-energy theory the spinon is a
mobile impurity in the Tomonaga-Luttinger liquid
(\ref{eq:Luttinger_with_spin}). It is described by the free-spinon
Hamiltonian (\ref{eq:H_0_spinon}), and the various perturbations
(\ref{eq:spinon_phonon_coupling}), (\ref{eq:Kondo_top}),
(\ref{eq:J_Kondo_prime}), (\ref{eq:V_rho_sigma}), and (\ref{eq:V_g}).

The perturbations  (\ref{eq:spinon_phonon_coupling}) and (\ref{eq:Kondo_top}) are nominally marginal, and the remaining ones are irrelevant. The perturbation (\ref{eq:spinon_phonon_coupling}) does not scale at all, whereas (\ref{eq:Kondo_top}) scales logarithmically at low energies. The scaling of the coupling constant $J_K$  can be understood by noticing that the
perturbation of the form (\ref{eq:Kondo_top}) describes the two-channel Kondo
problem.\cite{lamacraft_kondo_2008}

Positive  $J_K$ corresponds to the antiferromagnetic Kondo
problem. In that case, the coupling constant grows at low energies, and
the two-channel Kondo problem scales to an intermediate-coupling fixed
point,\cite{nozieres_kondo_1980} where the impurity spin is fully
screened.  The enhanced coupling suppresses the mobility of the
impurity\cite{lamacraft_kondo_2008} to $\mu\propto T^{-2}$, compared
to $\mu\propto T^{-4}$ in the spinless
case.\cite{castro_neto_dynamics_1996} As a result, the diffusion
constant in momentum space, $B=2T/\mu$,\cite{lifshitz_physical_1981}
should scale as $B\propto T^3$, in contrast to $B\propto T^5$ for a
spinless impurity.\cite{matveev_equilibration_2012}

Our interest in diffusion of mobile impurities is motivated by the
problem of scattering of spinon excitations in the Heisenberg chain
(\ref{eq:spin-chain}).  Their spins are not screened even at $T=0$,
indicating that the coupling constant $J_K$ may not be positive.
Negative $J_K$ corresponds to the ferromagnetic Kondo problem, in
which the coupling constant scales to zero logarithmically, and the
impurity spin remains unscreened.  This form of scaling of the
coupling of spinons to low-energy spin excitations in one-dimensional
Fermi systems was discussed in
Ref.~\onlinecite{schmidt_spin-charge_2010}.  At finite temperatures
the scaling suppresses the coupling constants by a factor of order
$\ln (J/T)$, which will not play an important role in our theory
compared to the much stronger temperature dependence of the
equilibration rate (\ref{eq:equilibration_rate_with_spin}).

\subsubsection{Scattering of spinons by charge excitations}
\label{sec:scatt-spinons-charge}

We start the evaluation of the diffusion constant $B$  by studying the contribution $B_{\rho\rho}$ originating from
the coupling of the spinon to the charge excitations of the Wigner
crystal. The latter are phonons discussed in Sec.~\ref{sec:phonons}.  The
spinon couples to phonons because its energy $\epsilon\sim J$ depends
on the electron density.  It
is important to note that such a coupling is insensitive to the spin
degree of freedom of the spinon.  In that sense, the problem is
equivalent to that of a spinless mobile impurity diffusing in a
Luttinger liquid.\cite{castro_neto_dynamics_1996} The diffusion
constant in momentum space was
expressed\cite{matveev_equilibration_2012} in terms of the spectrum of
the impurity and its dependence on density, and is given by
Eq.~(\ref{eq:chi_spinless}).  We thus obtain
$B_{\rho\rho}$ by substituting $\Delta=\pi J/2$ and $v=v_\rho$ into
the above expressions.  Because of the exponential dependence
(\ref{eq:J}) of $J$ on density, the first term in the parentheses in
Eq.~(\ref{eq:chi_spinless}) gives the dominant contribution, and we
find
\begin{equation}
  \label{eq:B_rho_rho}
  B_{\rho\rho}=\frac{\pi^5\eta^4}{240}
      \frac{T^5J^2}{\hbar^5 n^4 a_B^2 m^2 v_\rho^8}.
\end{equation}
The phonon velocity $v_\rho$ can be obtained from
Eq.~(\ref{eq:omega_q}).  At $q\to0$ one finds
\begin{equation}
  \label{eq:v_rho}
  v_\rho=\frac{\sqrt{2\mathcal L}}{\hbar n}
        \frac{e^2}{\varepsilon a_B}
        (na_B)^{3/2}.
\end{equation}
Theoretically, the parameter $\mathcal L$ diverges logarithmically at
$q\to 0$, see Appendix~\ref{sec:frequencies}.  On the other hand, the
scattering of the spinons is dominated by thermal phonons, for which
\begin{equation}
  \label{eq:logarithm}
  \mathcal L=\ln\frac{e^2(na_B)^{3/2}}{\varepsilon a_B T}.
\end{equation}
Alternatively, if the Coulomb interaction is screened by a metal gate
at a distance $d$ from the Wigner crystal, the divergence is cut off
as $\mathcal L=\ln(nd)$.

\subsubsection{Mixed scattering processes}
\label{sec:scatt-spin-charge}

We now consider the contribution $B_{\rho\sigma}$ to the diffusion
constant (\ref{eq:B_three_channels}).  This contribution accounts for
the scattering processes in which one of the two branches $a$ and $e$
belongs to the charge sector and the other to the spin sector.  The
first step is to obtain the corresponding term in the
$T$-matrix using the standard perturbative expression
\begin{equation}\label{eq:T_pertubative}
    \hat T=\hat V +\hat V \frac{1}{E_i - H_0} \hat V + \cdots ,
\end{equation}
where $E_i$ is the energy of the initial state. We are interested in
the on-shell matrix elements for which the energies in the initial and
final states are equal.  The relevant contribution has the form
\begin{equation}\label{eq:T_rho sigma}
    t_{m}n u'(Y)\,\bm S\cdot \left[\bm s^R(Y)
           + \bm s^L(Y)\right].
\end{equation}
Similar to Eq.~(\ref{eq:chi_spinless}), there are three contributions
to $t_m$ arising from different perturbations in the effective
Hamiltonian.

The simplest contribution is obtained by applying the perturbation
(\ref{eq:J_Kondo_prime}) in the first order,
\begin{equation}\label{eq:t_m_1}
t_m^{(1)}= -  J'_K .
\end{equation}
This contribution is analogous to the first term in
Eq.~(\ref{eq:chi_spinless}).  The other two contributions arise in the
second order perturbation theory.  Combining the perturbations $\hat
V_K$ and $\hat V_{\rho\sigma}$ given by Eqs.~(\ref{eq:Kondo_top}) and
(\ref{eq:V_rho_sigma}) in the second order we obtain
\begin{equation}\label{eq:t_m_2}
    t_{m}^{(2)}= J_K \frac{v_\sigma'}{v_\sigma}.
\end{equation}
This contribution is analogous to the second term in
Eq.~(\ref{eq:chi_spinless}). In this process $\hat V_K$ describes
scattering of a spinon off of a virtual spin boson, and $\hat
V_{\rho\sigma}$ describes the interaction of the latter with the charge
and spin bosons present in the initial and final states. The final
contribution arises in second order in the perturbations $\hat
V_{\mathrm {ph}}$ and $\hat V_K$ given by
Eqs.~(\ref{eq:spinon_phonon_coupling}) and (\ref{eq:Kondo_top}). It
describes the scattering process in which the spinon interacts
sequentially with the spin and charge bosons present in the initial
and final states. In this process, the absorption of the initial state
boson and the emission of the final state boson can happen in a different
order, which leads to a near cancellation of the corresponding
contributions. A finite result arises only due to the curvature of the
spinon spectrum and is given by
\begin{equation}\label{eq:t_m_3}
    t_m^{(3)}=\mathrm{sgn} (p_a p_e)\frac{\Delta ' J_K }{m^* v_\rho v_\sigma },
\end{equation}
where $m^*$ is defined in Eq.~(\ref{eq:Delta_and_mass}). This
contribution is analogous to the last term in
Eq.~(\ref{eq:chi_spinless}).

The full matrix element $t_m$ in Eq.~(\ref{eq:T_rho sigma}) is given
by the sum of the contributions (\ref{eq:t_m_1})--(\ref{eq:t_m_3}).
At low densities all three contributions are exponentially small. For
the Heisenberg model (\ref{eq:spin-chain}) the quantities $J_K$ and
$v_\sigma$ are controlled by a single parameter $J$.  As a result the
combined contribution
\begin{equation}\label{eq:t_m_12}
    t_m^{(1)} + t_m^{(2)}=J_K\left[\frac{v_\sigma' }{v_\sigma}- \frac{J_K'}{J_K} \right]
\end{equation}
vanishes. A nonvanishing result appears only if one takes into account
next nearest neighbor exchange coupling $\tilde J$.  Using
Eq.~(\ref{eq:tildeJ_large_rs}) we conclude
\begin{equation}\label{eq:t_m_12_estimate}
    t_m^{(1)} + t_m^{(2)}\propto \tilde J \propto  \exp\left( - \frac{2\eta + \Delta \eta}{\sqrt{n a_B}}\right).
\end{equation}
Noticing that $\Delta \sim n J_k \sim J$ and that $ m^* v_\sigma =
\hbar n$ we find
\begin{equation}\label{eq:t_m_3_estimate}
    t_m^{(3)} \propto J^2 \propto \exp\left( - \frac{2\eta}{\sqrt{n a_B}}\right).
\end{equation}
Keeping in mind that $\Delta \eta \approx -0.013$, we conclude that in
the limit of low density Eq.~(\ref{eq:t_m_12_estimate}) gives the
dominant contribution to $t_m$, but at $na_B < 10^4$ the difference
between the exponents in Eqs.~(\ref{eq:t_m_12_estimate}) and
(\ref{eq:t_m_3_estimate}) is insignificant.

In order to obtain the contribution $B_{\rho\sigma}$ to the diffusion
constant (\ref{eq:B_three_channels}) we substitute Eq.~(\ref{eq:T_rho
  sigma}) into Eq.~(\ref{eq:M_definition}) and obtain
\begin{eqnarray}
  \label{eq:M_alpha_beta_sigma_rho_result}
    M^{\alpha\beta}_{p_ap_e}(\delta p)&=& t_m^2
     (\delta_{\alpha,\beta}
            +2\delta_{\alpha,-\beta})
     \delta_{p_a-p_e,\delta p}
\nonumber\\
  &&\times\frac{ n |p_a p_e|}{8\pi \hbar m v_\rho L^2}
          N_{p_a}^{(a)}\big(N_{p_e}^{(e)}+1\big),
\end{eqnarray}
where $N_{p}^{(a,e)}$ denote either $N_p^{(\rho)}$ or $N_p^{(\sigma)}$, depending on the nature of the $a$ and $e$ branches, and are given by
Eq.~(\ref{eq:Bose_distribution}) with $u=0$. Then, using
Eqs.~(\ref{eq:B_definition})--(\ref{eq:W^ae}), we find
\begin{equation}\label{eq:B_rho_sigma}
    B_{\rho\sigma}=t_m^2 \frac{2\pi^2 nT^5}{5\hbar^6 m v_\rho^3 v_\sigma^4}.
\end{equation}
In analogy with the result (\ref{eq:B_rho_rho}) for $B_{\rho\rho}$,
the contribution (\ref{eq:B_rho_sigma}) scales as $T^5$.  On the other
hand, the exponentially small velocity of the spin excitations
$v_\sigma$ in the denominator of Eq.~(\ref{eq:B_rho_sigma}) enhances
$B_{\rho\sigma}$ as compared to $B_{\rho\rho}$,
\begin{equation}
\label{eq:B_rho_rho_B_rho_sigma}
  \frac{B_{\rho\rho}}{B_{\rho\sigma}} \propto \exp\left(-\frac{ 2\eta}{\sqrt{na_B}}\right),
\end{equation}
where we used the estimate (\ref{eq:t_m_3_estimate}) for $t_m$.  From
Eq.~(\ref{eq:B_rho_rho_B_rho_sigma}) we conclude that at low electron
density $B_{\rho\sigma}\gg B_{\rho\rho}$.

\subsubsection{Scattering of spinons by spin excitations}
\label{sec:scatt-spin-spin}

We now turn to the contribution $B_{\sigma\sigma}$ to the diffusion
constant (\ref{eq:B_three_channels}).  It arises from the scattering
processes in which both the $a$ and $e$ branches belong to the spin
sector.  The leading contribution to the $T$-matrix appears in the
second order with perturbations (\ref{eq:Kondo_top}) and
(\ref{eq:H_g}).  For a spinon near the top of the spectrum the
on-shell part is given by
\begin{equation}
  \label{eq:T-matrix_spin-spin}
  -\frac{i}{L}J_K(J_K+2\pi\hbar g v_\sigma)
   \sum_p\frac{e^{2ipY/\hbar}}{v_\sigma p}\bm S\cdot[\bm s^R(p)\times \bm s^L(p)],
\end{equation}
where we have introduced the Fourier transforms of the spin density
operators via
\begin{equation}
  \label{eq:spin_density_Fourier}
  \bm s^{R,L}(Y)=\frac{1}{\sqrt L}\sum_p\bm s^{R,L}(p)\, e^{ipY/\hbar}.
\end{equation}
It is important to note that the scattering matrix elements are
enhanced at small momenta $p$ by the denominator in
Eq.~(\ref{eq:T-matrix_spin-spin}).  Similar contributions proportional
to $1/p$ in the second-order calculations for the $\rho\rho$ and
$\rho\sigma$ channels cancel each other.  The absence of such a
cancelation in the $\sigma\sigma$ channel is due to the
noncommutativity of the spin operators in the perturbations
(\ref{eq:Kondo_top}) and (\ref{eq:H_g}).  The enhancement of
quasiparticle scattering in the presence of spins was first pointed
out in Ref.~\onlinecite{karzig_energy_2010}.  It is worth noting that
the nonlocal nature of Eq.~(\ref{eq:T-matrix_spin-spin}) precludes the
possibility of such terms appearing as perturbations in the
Hamiltonian of a spinon interacting with the Tomonaga-Luttinger
liquid.  We therefore do not expect first order contributions to the
$T$-matrix of the form (\ref{eq:T-matrix_spin-spin}).

Combining Eqs.~(\ref{eq:T-matrix_spin-spin}) and
(\ref{eq:B_definition})-(\ref{eq:M_definition}), we obtain the
contribution to the diffusion constant in the form
\begin{equation}
  \label{eq:B_sigma_sigma_result}
  B_{\sigma\sigma}=\frac{J_K^2(J_K+2\pi\hbar g v_\sigma)^2 T^3}
                     {8\pi\hbar^7v_\sigma^6}.
\end{equation}
The aforementioned enhancement of the scattering in the spin-spin
channel results in $B_{\sigma\sigma}\propto T^3$, compared to the $T^5$
dependence of $B_{\rho\rho}$ and $B_{\rho\sigma}$. An analogous
dependence $B\propto T^3$ was recently predicted in the limit of
weakly interacting electrons.\cite{Rieder-unpublished}

At strong interactions, the magnitude of $B_{\sigma\sigma}$ is
controlled by the coupling constants $J_K$ and $v_\sigma$.  In a
Heisenberg chain (\ref{eq:spin-chain}) there is only a single energy
scale, $J$.  Thus, one expects $J_K\sim -gJ/n$, where the logarithmic
factor $g=1/\ln(J/T)$ appears as a result of the usual renormalization
of the ferromagnetic Kondo coupling constant.  Taking into account the
expression (\ref{eq:v_sigma}) for $v_\sigma$, we conclude that the two
terms in the combination $J_K+2\pi\hbar g v_\sigma$ in
Eq.~(\ref{eq:B_sigma_sigma_result}) are of the same order of magnitude, and of opposite signs.  On the other hand, the integrability of the
Heisenberg model precludes real scattering processes, i.e., the above
combination of the coupling constants must vanish.

Nonvanishing scattering appears due to perturbations that break the
integrability of the spin chain (\ref{eq:spin-chain}).  The simplest
such perturbation is the next-nearest neighbor coupling $\tilde J$.
Its presence results in $J_K+2\pi\hbar g v_\sigma\propto \tilde J/n$.
Substituting this estimate into Eq.~(\ref{eq:B_sigma_sigma_result})
and omitting the logarithmic factors, we obtain
\begin{equation}
  \label{eq:B_sigma_sigma_estimate}
  B_{\sigma\sigma}\sim \frac{n^2}{\hbar}\,\frac{\tilde J^2 T^3}{J^4}.
\end{equation}
This estimate applies at $T\lesssim J$.  Comparing
Eq.~(\ref{eq:B_sigma_sigma_estimate}) with the estimate of
$B_{\rho\sigma}$ given by Eqs.~(\ref{eq:t_m_12_estimate}) and
(\ref{eq:B_rho_sigma}) we find
\begin{equation}
  \label{eq:B_rho_sigma_B_sigma_sigma}
    \frac{B_{\rho\sigma}}{B_{\sigma\sigma}}
    \propto \exp\left(-\frac{ 2\eta}{\sqrt{na_B}}\right)
\end{equation}
at $T\sim J$; the ratio is even lower at $T\ll J$.

In addition to the next-nearest neighbor coupling in the spin chain,
the integrability of the problem is also broken by the spin-charge
coupling.  The leading contribution to $B_{\sigma\sigma}$ in this
mechanism is obtained in second order in perturbations
(\ref{eq:spinon_phonon_coupling}) and (\ref{eq:V_g}), whereby a spinon
is coupled to acoustic spin excitations via an exchange of a virtual
phonon.  Instead of $\tilde J$, such contributions to
$B_{\sigma\sigma}$ contain $J^2$, and result in essentially the same
estimate (\ref{eq:B_rho_sigma_B_sigma_sigma}).  [See an analogous
discussion below Eq. (\ref{eq:t_m_3_estimate}).]  However, the
processes of coupling by a virtual phonon do not involve noncommuting
spin operators, and thus lack the enhancement of the scattering due to
the small momentum in the denominator of
Eq.~(\ref{eq:T-matrix_spin-spin}).  As a result, their contributions
to $B_{\sigma\sigma}$ scale as $T^5$, and are small compared to
Eq.~(\ref{eq:B_sigma_sigma_result}) at $T\ll J$.  We therefore
conclude that among the three terms in the diffusion coefficient
(\ref{eq:B_three_channels}), the spin channel contribution
(\ref{eq:B_sigma_sigma_result}) always dominates.

\section{Summary and Discussion}
\label{sec:summary}

The results obtained in this paper enable us to obtain the temperature
dependent correction to the conductance of quantum wires at strong
interactions in the spin-degenerate case.  Previously such corrections
at strong interactions were studied only for spin-polarized
electrons.\cite{matveev_equilibration_2010,
  matveev_equilibration_2011} The spin-degenerate case represents a
significantly more complicated problem, whose treatment requires
consideration of all perturbations breaking integrability of the
model, carried out above.  We start by summarizing our results.

In Sec.~\ref{sec:corrections} we identified three types of
perturbations to the Hamiltonian (\ref{eq:H_rho}),
(\ref{eq:spin-chain}) of the one-dimensional Wigner crystal.  They
include the anharmonic corrections (\ref{eq:perturbation}) in the
charge sector, the next-nearest neighbor exchange of the spins
(\ref{eq:three-particle_perturbation}), and the coupling of the charge
and spin degrees of freedom (\ref{eq:v_rho_sigma_1}).  The magnitude
of the next-nearest neighbor exchange was obtained by numerical
treatment of the WKB action in Sec.~\ref{sec:WKB}.

All of the above perturbations break integrability of the Hamiltonian
(\ref{eq:H_rho}), (\ref{eq:spin-chain}), and thus enable scattering of
quasiparticle excitations off each other.  In
Sec.~\ref{sec:rate-full-equil} we applied these results to the
evaluation of the rate of full equilibration of the one-dimensional
Wigner crystal.  We have found that the dominant process of
equilibration involves scattering of a high-energy spinon by two
acoustic spin excitations.  The resulting equilibration rate is given
by Eq.~(\ref{eq:equilibration_rate_with_spin}), with the dominant
contribution to the spinon diffusion constant $B$ given by
Eq.~(\ref{eq:B_sigma_sigma_result}).  Using
Eq.~(\ref{eq:B_sigma_sigma_estimate}) one estimates the equilibration
rate as
\begin{equation}
  \label{eq:equlibration_rate_estimate}
  \tau^{-1}\sim \frac{\tilde J^2}{\hbar\sqrt{JT}}
               \exp
                 \left(
                   - \frac{\pi J}{2T}
                 \right).
\end{equation}
The magnitude $\tilde J$ of the next-nearest neighbor coupling is
given by Eq.~(\ref{eq:tilde_J_integral}).

As one can see from Eq.~(\ref{eq:full_momentum}), conservation of the
total momentum of the electron liquid means that the full
equilibration is accompanied by backscattering of electrons.  This
enables one to relate the equilibration rate to the conductance of
long uniform quantum wires.\cite{micklitz_transport_2010,
  matveev_equilibration_2011} For example, in the limit of strong
interactions, the interaction induced correction to the quantum
conductance $2e^2/h$ is given by
\begin{equation}
  \label{eq:conductance-strong-interactions}
  \delta G=-\frac{2e^2}{h}\frac{8\hbar nT^2}{3\pi^3J^3}\,
     \frac{L}{\tau}.
\end{equation}
The correction grows with the length of the wire $L$ and eventually
saturates at $L\sim J\tau/\hbar n$.\cite{matveev_equilibration_2011}
In the shorter wires the correction to the quantized conductance is
proportional to the equilibration rate $\tau^{-1}$.  Combining
Eqs.~(\ref{eq:equlibration_rate_estimate}) and
(\ref{eq:conductance-strong-interactions}) one then obtains $\delta
G\propto T^{3/2}\exp(-\pi J/2T)$.  The activated behavior of the
correction to conductance of quantum point contacts was observed
experimentally.\cite{kristensen_bias_2000} The activation temperature
reported in Ref.~\onlinecite{kristensen_bias_2000} was rather small,
$T_A\sim 1$K, and grew rapidly with electron density $n$.  These
observations are consistent with the fact that the exchange constant
$J$ given by Eq.~(\ref{eq:J}) is exponentially small at $na_B\ll1$.

\acknowledgments The authors are grateful to L. I. Glazman and
B. I. Halperin for stimulating discussions.  Work at Argonne National
Laboratory was supported by the U.S. Department of Energy, Office of
Science, Materials Sciences and Engineering Division. Work at the
University of Washington was supported by U. S. Department of Energy
Office of Science, Basic Energy Sciences under award number
DE-FG02-07ER46452.

\appendix

\section{Evaluation of phonon frequencies in the Wigner crystal at low
wavenumber}
\label{sec:frequencies}

The frequencies of phonons in a one-dimensional Wigner crystal are
given by Eq.~(\ref{eq:omega_q}).  Upon the introduction of the
dimensionless time $\tau$ via Eq.~(\ref{eq:rescaled_variables}), the
phonon frequencies become
\begin{equation}
  \label{eq:frequencies_exact_dimensionless}
  \omega_q^2=4 \sum_{l=1}^\infty \frac{1}{l^3}[1-\cos(ql)].
\end{equation}
At $q\to 0$ one can expand the cosine and obtain the logarithmic behavior
\begin{equation}
  \label{eq:frequencies_small_q}
  \omega_q^2=2q^2\ln\frac{\chi}{q}.
\end{equation}
A more careful calculation is required to obtain the value of the
constant $\chi$, which is the subject of this Appendix.

We start by substituting the identity
\[
\frac{1}{l^3}=\frac12\int_0^\infty x^2 e^{-lx} dx
\]
into Eq.~(\ref{eq:frequencies_exact_dimensionless}) and performing the
trivial summation over $l$ in the resulting expression:
\begin{equation}
  \label{eq:omega_rewritten}
\omega_q^2=\int_0^\infty x^2
           \left[
             \frac{2}{e^x -1}-\frac{1}{e^{x+iq} -1}-\frac{1}{e^{x-iq} -1}
           \right]dx.
\end{equation}
Let us now split the above integral into two, with the first one taken
from 0 to $x_0$ and the second from $x_0$ to $\infty$.  The value of
$x_0$ is chosen such that $q\ll x_0\ll 1$.  In the first integral one
can expand $e^x\simeq 1+x$ and obtain
\begin{eqnarray*}
&&\int_0^{x_0} x^2
         \left[
             \frac{2}{e^x -1}-\frac{1}{e^{x+iq} -1}-\frac{1}{e^{x-iq} -1}
           \right]dx\\
&&\simeq
2q^2\int_0^{x_0} \frac{x\,dx}{x^2+q^2}\simeq 2q^2\ln\frac{x_0}{q}.
\end{eqnarray*}
To evaluate the second integral we expand the integrand to second
order in small $q$ and obtain
\begin{eqnarray*}
&&\int_{x_0}^\infty x^2
           \left[
             \frac{2}{e^x -1}-\frac{1}{e^{x+iq} -1}-\frac{1}{e^{x-iq} -1}
           \right]dx\\
&&\simeq
\int_{x_0}^\infty x^2 q^2\left(\frac{1}{e^x-1}\right)'' dx
\simeq
q^2[-2\ln x_0 +3].
\end{eqnarray*}
The total integral in Eq.~(\ref{eq:omega_rewritten}) is independent of
$x_0$ and given by
\[
  \omega_q^2=2q^2\ln\frac1q +3q^2.
\]
Comparing this result with Eq.~(\ref{eq:frequencies_small_q}) we
obtain $\chi=e^{3/2}\approx 4.48$.

\section{Analytical treatment of the instanton action}
\label{sec:analytical-treatment}

The instanton action (\ref{eq:eta_definition}) is minimized for the
configuration in which the electrons rest at the positions of static
equilibrium $X_l(\tau)=l$.  Small fluctuations near the minimum can be
studied by introducing the displacements $u_l$ of electrons from
equilibrium positions
\begin{equation}
  \label{eq:u_definition}
  X_l(\tau) = l + u_l(\tau).
\end{equation}
Substituting Eq.~(\ref{eq:u_definition}) into
(\ref{eq:eta_definition}) and expanding in $u_l$ one finds the
quadratic action
\begin{equation}
  \label{eq:eta_quadratic}
  \eta^{(2)}=\int_{-\infty}^\infty
     \left(
       \sum_l \frac{1}{2}\, \dot u_l^2
       +\sum_{l<l'} \frac{(u_{l'}-u_l)^2}{(l'-l)^3}
     \right)d\tau.
\end{equation}
Upon the Fourier transformation of the displacements
\begin{equation}
  \label{eq:u_Fourier}
  u_l(\tau)=\int\frac{dqd\omega}{(2\pi)^2}\,
            e^{iql-i\omega \tau} u_{q\omega}
\end{equation}
the quadratic action takes the form
\begin{equation}
  \label{eq:eta_quadratic_Fourier}
  \eta^{(2)}=\int\frac{dqd\omega}{(2\pi)^2}\,
            \frac{1}{2}(\omega^2+\omega_q^2)|u_{q\omega}|^2.
\end{equation}
The phonon frequencies $\omega_q$ are evaluated in
Appendix~\ref{sec:frequencies}.

\subsection{Single instanton}
\label{sec:single-instanton}

The amplitude of the nearest-neighbor exchange between sites 0 and 1
is determined by the action of an instanton with the boundary
conditions $X_0(-\infty)=X_1(+\infty)=0$,
$X_0(+\infty)=X_1(-\infty)=1$, and $X_l(\pm\infty)=l$ for all $l\neq
0,1$.  Minimization of Eq.~(\ref{eq:eta_definition}) over all
$X_l(\tau)$ then results in the single-instanton action $\eta\approx
2.80$.\cite{klironomos_exchange_2005}

It is instructive to study interaction of the instanton with
long-wavelength fluctuations of the displacements $u_l(\tau)$.  A
shift of all $u_l$ by a constant $\delta u$ corresponds to the
translation of the whole crystal and has no effect on the action.  A
uniform $\dot u_l=v$ has the meaning of the velocity of the crystal.
Although the instanton action is affected by the motion of the system
as a whole, the effect should be even in $v$, and thus instanton
action does not couple to $\dot u_l$ in first order.  On the other
hand, the spatial derivative $\partial_l u$ corresponds to stretching
the crystal and results in the change of density
\begin{equation}
  \label{eq:density}
  n=\frac{n_0}{1+\partial_l u}.
\end{equation}
Since the dimensionless action (\ref{eq:eta_definition}) assumes that
density equals 1, the effect of the change $n_0\to n$ should be
obtained from the full expression (\ref{eq:action_rescaled}).
Substituting Eq.~(\ref{eq:density}) one then obtains
\[
\frac1\hbar S_1=\frac{\eta}{\sqrt{n_0 a_B}}\sqrt{1+\partial_l u}
\simeq\frac{1}{\sqrt{n_0 a_B}}
      \left(\eta+\frac{\eta}{2}\partial_l u\right).
\]
Thus to lowest order the instanton couples linearly to fluctuations of
the field $u_l(\tau)$,
\begin{equation}
  \label{eq:instanton_coupling}
  \delta\eta=d\,\partial_lu,
  \quad
  d=\frac{\eta}{2},
\end{equation}
where the derivative $\partial_lu$ is taken at $l$ and $\tau$
corresponding to the location of the instanton.

\subsection{Interaction of instantons}
\label{sec:inter-inst}

Let us now consider the configuration of $X_l(\tau)$ corresponding to
two instantons at positions $(0,0)$ and $(l,\tau)$.  Assuming the
instantons are far from each other, $l^2+\tau^2\gg 1$, the action can
be presented as
\begin{equation}
  \label{eq:two_instantons}
  \eta_2(l,\tau)=2\eta+\delta\eta(l,\tau),
\end{equation}
where the small correction $\delta\eta(l,\tau)$ has the meaning of the
interaction between the instantons.  To find it one can minimize the
quadratic action (\ref{eq:eta_quadratic}) with the perturbation
$d[\partial_l u(0,0)+\partial_l u(l,\tau)]$.  An alternative approach
is to find the shape $u(l,\tau)$ of the instanton centered at $(0,0)$
at large distance using the perturbation $d\partial_l u(0,0)$ and then
apply Eq.~(\ref{eq:instanton_coupling}) to find coupling to the second
instanton.  Following this approach one easily obtains
\begin{equation}
  \label{eq:instanton_tail}
  u(l,\tau)=d\int\frac{dqd\omega}{(2\pi)^2}\,
            \frac{iqe^{iql-i\omega\tau}}{\omega^2+\omega_q^2}
\end{equation}
at large distances from the first instanton.  The interaction of the
instantons is obtained by differentiating the above expression:
\begin{eqnarray}
  \label{eq:instanton_interaction}
  \delta\eta(l,\tau)&=&-d^2\int\frac{dqd\omega}{(2\pi)^2}\,
            \frac{q^2e^{iql-i\omega\tau}}{\omega^2+\omega_q^2}
\nonumber\\
  &=&-d^2\int\frac{dq}{4\pi}\,
            \frac{q^2}{\omega_q}\,e^{iql-\omega_q|\tau|}.
\end{eqnarray}

\subsection{Electrostatic analogy}
\label{sec:electr-anal}

Because of the logarithmic singularity in $\omega_q$, see
Eq.~(\ref{eq:frequencies_small_q}), the remaining integral in
Eq.~(\ref{eq:instanton_interaction}) cannot be easily performed.  On
the other hand, considerable progress can be made by replacing
$\ln(\chi/q)$ with a constant, $\omega_q\simeq sq$.  Using this
approximation, we immediately find
\begin{equation}
  \label{eq:instanton_interaction_electrostatic}
  \delta\eta(l,\tau)=\frac{d^2}{2\pi s}\frac{l^2-s^2\tau^2}{(l^2+s^2\tau^2)^2}.
\end{equation}
This result has the form of interaction of two dipoles in
two-dimensional space $(l,s\tau)$.

The analogy is developed as follows.  The action
(\ref{eq:eta_quadratic}) is presented as
\begin{equation}
  \label{eq:eta_electrostatic}
  \eta_{el}=\int d^2r \left[\frac{s}{2} (\nabla u)^2 -\rho u\right],
\end{equation}
where we have added the ``charge density'' term $\rho u$.
Minimization of $\eta_{el}$ with respect to $u$ gives the Poisson
equation $\nabla^2u=-\rho/s$, with $s$ playing the role of
$\varepsilon_0$ in SI units.  Substitution of $\rho=-s\nabla^2u$ into
Eq.~(\ref{eq:eta_electrostatic}) enables one to express the action in
terms of the ``electric potential'' $u$,
\begin{equation}
  \label{eq:energy_electric_field}
  \eta_{el}=-\int d^2r \frac{s}{2} (\nabla u)^2.
\end{equation}
Thus the action $\eta_{el}$ is given by the energy of the effective
electric field with the opposite sign.

The coupling term (\ref{eq:instanton_coupling}) corresponds to
$\rho=d\,\partial_l\delta({\bf r})$, analogous to the charge
distribution in a dipole pointing in the $l$ direction.  In two
dimensions, the interaction between two such dipoles is given by the
negative of Eq.~(\ref{eq:instanton_interaction_electrostatic}), as
expected in our electrostatic analogy.

\bibliographystyle{apsrev}

\end{document}